\documentclass[journal]{IEEEtran}

\usepackage{booktabs}
\usepackage{multirow}
\usepackage{threeparttable}   
\usepackage{graphicx}         
\usepackage{amsmath}
\usepackage{amssymb}
\usepackage{amsthm}
\usepackage{xcolor}
\definecolor{donegreen}{RGB}{0,140,70}
\definecolor{pendingpurple}{RGB}{140,70,180}
\usepackage{tabularx}
\usepackage{makecell}
\usepackage{pifont}
\usepackage{float}
\usepackage{dblfloatfix}
\usepackage{cuted}
\usepackage{booktabs}
\usepackage{capt-of}
\usepackage{array}
\usepackage[ruled,vlined,linesnumbered]{algorithm2e}
\usepackage{xcolor}
\usepackage{subcaption}
\usepackage[caption=false,font=footnotesize]{subfig}
\usepackage{capt-of}
\usepackage{caption}
\usepackage{array}

\newcommand{\cmark}{\ding{51}}
\newcommand{\xmark}{\ding{55}}

\begin{document}

\bstctlcite{IEEEexample:BSTcontrol}


\title{Multi-Timescale Context-Aware Radio Resource Optimization for Harmonious NTN-TN Coexistence: A Digital Twin and Open RAN Enabled Framework}


\author{Ecem Izde Kaya,
Md. Zoheb Hassan,~\IEEEmembership{Member,~IEEE},
and Georges Kaddoum,~\IEEEmembership{Senior Member,~IEEE}%
\thanks{ E. I. Kaya and G. Kaddoum are with the $\acute{\text{E}}$cole de technologie sup$\acute{\text{e}}$rieure (ETS), Université du Qu\'ebec, Montr\'eal, QC, Canada. M. Z. Hassan is with the  Universit\'e Laval, Qu\'ebec City, QC, Canada. (Corresponding author's e-mail: md-zoheb.hassan@gel.ulaval.ca).}
\thanks{This work has been submitted to the IEEE for possible publication. Copyright may be transferred without notice, after which this version may no longer be accessible.}
}
\maketitle

\begin{abstract}
Coexistence between non-terrestrial networks (NTNs) and terrestrial networks (TNs) will be essential for future wireless systems. While spectrum sharing with TNs will enable NTNs to more efficiently expand coverage and use resources, a key challenge that remains is interference. In particular, terrestrial downlink transmissions can generate harmful aggregate interference towards low-Earth orbit (LEO) satellite receivers, making interference-aware resource management essential for coexistence.
To address this need, in this article, we present DT-STABLE, an Open Radio Access Network (O-RAN)-enabled, digital-twin (DT)-assisted framework for context-aware NTN–TN coexistence. 
By using site-specific contextual information, the DT estimates interference perceived at the LEO satellites, enabling proactive interference management without extensive real-time measurement overhead. We formulate a joint beam activation and user association problem accounting for RF-chain limits, load balancing, and aggregate satellite interference constraints, prove its NP-hardness, and develop an efficient DT-guided algorithm for practical scheduling. DT-STABLE executes a multi-timescale optimization by coupling short-timescale radio resource adaptation with periodic mid- and long-timescale updates of the DT's predictive state. This synchronization is maintained through a continuous two-way exchange of context, measurements, and control information over standardized O-RAN interfaces, ensuring that the twin remains accurate and aligned with the physical network. The results of site-specific simulations under worst-case satellite geometry at 28 GHz band, with LEO satellites at low elevation angles, demonstrate that DT-STABLE outperforms existing approaches in balancing rate and interference utility, achieving 89.63\% interference-to-noise ratio (INR) satisfaction and a mean sum rate of 210.96~bps/Hz, while attaining nearly 100\% INR satisfaction when satellite protection is prioritized. The proposed framework also reduces beam measurement overhead by 99.71\% as compared with exhaustive sweeping and achieves up to 33.1$\times$ lower runtime than the Gurobi-based exact optimization benchmark.
\end{abstract}

\vspace{-0.3cm}
\section{Introduction}
The integration of non-terrestrial networks (NTNs) with terrestrial networks (TNs) has been recognized by the International Telecommunication Union (ITU) and the Third Generation Partnership Project (3GPP) as a fundamental component of future 5G and 6G ecosystems \cite{3gpp_tr38811}. However, while this integration has a significant potential in terms of enhancing global connectivity, spectrum scarcity poses a critical challenge. 

As conventional sub-6 GHz bands become increasingly congested \cite{samsung_6g_spectrum_2022}, cellular systems have extended into frequency range 2 (FR2), spanning 24.25--52.6~GHz \cite{3gpp_ts38104}. At the same time, portions of this spectrum overlap with Ka-band frequencies used by satellite communication systems. In particular, the Federal Communications Commission (FCC) has authorized SpaceX's Gen2 Starlink non-geostationary orbit (NGSO) constellation for Earth-to-space operations in the 27.5--29.1~GHz and 29.5--30.0~GHz bands \cite{FCC2022SpaceXGen2}. This spectral overlap creates a coexistence challenge, as terrestrial downlink transmissions may cause harmful interference to satellite uplinks and degrade satellite performance~\cite{yastrebova2020theoretical}.

While 5G base stations (BSs) are primarily designed to serve terrestrial users, practical antenna patterns can radiate part of the transmitted power towards the sky via side lobes, highlighting the need for coordinated spectrum sharing solutions. Narrow and highly directional beams at millimetre-wave (mmWave) frequencies can improve spatial isolation by steering energy toward the intended receiver and limiting emissions in other directions~\cite{miller2021expanding}. However, interference remains a critical challenge for the reasons outlined below.

\par\noindent
\hangindent=0.9em\hangafter=1
$\bullet$\hspace{0.35em} Even with narrow beams, there are sidelobes, and sidelobe suppression broadens the main lobe \cite{hansen1992array,3gpp_tr38901}.

\par\noindent
\hangindent=0.9em\hangafter=1
$\bullet$\hspace{0.35em} Satellite interference can be significantly affected not only by the direct emissions of terrestrial transmitters, but also by the upscattering of terrestrial signals by surrounding objects such as buildings, trees, and vehicles, which further increases the aggregate interference observed at the satellite receiver \cite{molisch2025newspectrum}.

\par\noindent
\hangindent=0.9em\hangafter=1
$\bullet$\hspace{0.35em} Aggregate interference remains a significant concern, as true coexistence is determined by the overall antenna pattern and network deployment strategy, rather than by the interference characteristics of a single beam and single BS.

In recent years, the integration of open radio access network (O-RAN) architectures with advanced resource-management methods has gained increasing attention due to its ability to support flexible, scalable, and cost-efficient network operation \cite{seid2025multi}. O-RAN naturally supports these functions through programmable control loops, standardized interfaces, and AI/ML-oriented design, all of which enable virtual network intelligence to interact with physical RAN operation \cite{Alam2026ORAN}. Digital twin networks (DTNs) are virtual real-time replicas of physical networks supporting network design, diagnosis, simulation, what-if analysis, and AI/ML-driven optimization and control \cite{10148936}.The integration of digital twins (DTs) with O-RAN can improve spectrum management, enhance resource allocation, and support control-policy validation in a virtual environment before physical-network deployment \cite{Seth2026POST}.

\begin{table*}[!t]
\caption{Summary of previous studies on resource and interference management in integrated NTN--TN systems.}
\label{tab:related_works_summary}
\centering
\scriptsize
\setlength{\tabcolsep}{4pt}
\renewcommand{\arraystretch}{1.2}
\begin{tabularx}{\textwidth}{|>{\centering\arraybackslash}p{1.6cm}|>{\centering\arraybackslash}p{1.4cm}|>{\centering\arraybackslash}p{1.2cm}|>{\centering\arraybackslash}p{1.5cm}|>{\raggedright\arraybackslash}X|>{\raggedright\arraybackslash}X|}
\hline
\textbf{Reference} &
\makecell{\textbf{Interference}\\\textbf{Management}} &
\makecell{\textbf{FR2}\\\textbf{band}} &
\makecell{\textbf{Environment}\\\textbf{Awareness}} &
\textbf{Resource Variables} &
\textbf{Objective} \\
\hline
\cite{Almekhlafi2025Access} & \xmark & \xmark & \cmark & Satellite visibility, roaming & Access inequality analysis \\
\hline
\cite{abbasi2024channel} & \xmark & \xmark & \xmark & CSI prediction & Mitigate channel aging \\
\hline
\cite{peng2025dynamic} & \xmark & \xmark & \xmark & Scheduling, power, bandwidth & Maximize sum rate \\
\hline
\cite{Ozturk2026Sustainable} & \xmark & \xmark & \xmark & Cell switching, offloading & Minimize power consumption \\
\hline
\cite{rahman2025joint} & \cmark & \xmark & \xmark & Offloading, power, RBG & Maximize spectral efficiency \\
\hline
\cite{alam2024optimizing} & \xmark & \xmark & \xmark & Bandwidth, association, power & Maximize SLT, minimize energy \\
\hline
\cite{rahman2023coexistence} & \cmark & \cmark & \xmark & Exclusion-zone radii, frequency allocation & Maximize terrestrial data rates \\
\hline
\cite{kang2024terrestrial} & \cmark & \xmark & \xmark & Beamforming, spatial nulling & Mitigate uplink interference \\
\hline
\cite{tsampazi2026satellite} & \cmark & \xmark & \xmark & Power control, beamforming & Maximize power under INR/rate limits \\
\hline
\cite{jia2025joint} & \cmark & \xmark & \xmark & Beamforming, channel estimation & Maximize TN quality, limit interference \\
\hline
\cite{Zhao2025BeamHopping} & \xmark & \xmark & \cmark & Beam hopping, power allocation & Maximize throughput, fairness \\
\hline
\cite{Allouch2026} & \cmark & \xmark & \xmark &
Beamforming &
Maximize TN sum rate under INR constraints \\
\hline
\textbf{DT-STABLE} & \cmark & \cmark & \cmark & Beam selection, association, scheduling & Maximize capacity under INR constraints \\
\hline
\end{tabularx}
\vspace{-0.5cm}
\end{table*}



In this paper, we propose DT-STABLE, an O-RAN-enabled DT-assisted multi-time-scale radio resource management (RRM) framework for dynamic TN--NTN spectrum coexistence. Unlike conventional beam management strategies that rely on frequent real-time measurements, DT-STABLE uses a site-specific channel knowledge map (CKM), which provides location-specific channel knowledge, reflecting the local wireless propagation environment  \cite{10430216}. The CKM is generated offline from 3D environmental geometry, material information, and RF propagation modeling of a high-fidelity DT to predict terrestrial link quality and satellite interference exposure. This enables near real-time (near-RT), proactive beam activation and user scheduling under satellite visibility, aggregate interference, RF-chain, and beam-load constraints. DT-STABLE employs O-RAN closed-loop control to detect CKM prediction mismatches in dynamic wireless environments and to selectively trigger low-overhead beam sweeping for DT refinement over multi-time scale horizon. Major contributions can be summarized as follows:
\par\noindent
\hangindent=0.9em\hangafter=1
$\bullet$\hspace{0.35em} We formulate the NTN--TN spectrum coexistence problem as a constrained joint beam-activation and user-association problem that balances terrestrial service utility and satellite protection through a soft aggregate-interference penalty. We further show that the resulting problem is NP-hard.

\par\noindent
\hangindent=0.9em\hangafter=1
$\bullet$\hspace{0.35em} We develop DT-STABLE, an O-RAN-enabled
\textbf{D}igital \textbf{T}win
\textbf{S}atellite-\textbf{T}hreshold-\textbf{A}ware
\textbf{B}eam--\textbf{L}oad \textbf{E}ngine, where DT-in-the-loop optimization is built on top of the O-RAN architecture to support multi-timescale interference-aware terrestrial network control. The proposed framework jointly integrates long-term site-specific CKM construction, mid-term CKM refinement, and short-term beam--user scheduling across multiple control time scales.

\par\noindent
\hangindent=0.9em\hangafter=1
$\bullet$\hspace{0.35em} By using a multi-domain context-aware DT (CDT), we develop a 3D geo-environment and material-aware CKM that captures both terrestrial downlink channels and satellite interference, and facilitates proactive beam selection without exhaustive measurements. The terrestrial CKM is constructed using high-fidelity site-specific ray tracing (RT) with NVIDIA Sionna RT~\cite{hoydis2023sionnart}, incorporating real-world 3D building models, actual base station (BS) locations, and beamforming gains. The non-terrestrial CKM uses the OpenNTN channel model~\cite{Due2025OpenNTN}, incorporating two-line element (TLE) data-based satellite positions, satellite and BS antenna gains, stochastic channel variations, and propagation losses due to shadowing, clutter, weather, and atmospheric conditions.

\par\noindent
\hangindent=0.9em\hangafter=1
$\bullet$\hspace{0.35em} We introduce a low-overhead, periodic, and adaptive refinement mechanism that uses limited beam sweeping to update outdated or mismatched terrestrial CKM predictions, thereby enhancing robustness against DT modeling uncertainty.

\par\noindent
\hangindent=0.9em\hangafter=1
$\bullet$\hspace{0.35em} Using comprehensive site-specific simulations over the 28 GHz band,  we showcase DT-STABLE's efficacy for practical NTN-TN (uplink-downlink) coexistence scenarios while covering site-specific propagation, satellite visibility, and benchmark comparisons; we also evaluate its robustness under imperfect DT knowledge.

The remainder of this article is organized as follows. A summary of related works is provided in Section \ref{sec:related_work}. Section \ref{sec:system_model}
presents an overview of the system considered. In Section \ref{sec:dt_stable}, we present the DT-STABLE framework. The formulation of the optimization problem, the algorithm, and the multi-time scale resource optimization framework
are described in Section V. 
Finally, the performance evaluation and conclusion
are presented in Sections \ref{sec:performance_eval} and \ref{sec:conclusion}, respectively.

\vspace{-0.3cm}
\section{Related Works}
\label{sec:related_work}
\textit{Resource Management for NTN-TN:}
Unlike cellular networks, NTN systems face dynamic channel and topology conditions from satellite mobility, long propagation distances, and time-varying visibility, which motivates dedicated resource-management strategies. In low-Earth orbit (LEO) systems, long propagation delays and rapid channel variation can render channel state information (CSI) outdated, thus requiring hybrid beamforming and resource allocation based on channel-aging \cite{abbasi2024channel}. Time-varying satellite visibility creates heterogeneous service conditions, with remote-region users experiencing fewer visible satellites, lower elevation angles, and longer service discontinuities than urban mid-latitude users \cite{Almekhlafi2025Access}. By contrast, simultaneous multi-satellite visibility enables cooperative transmission and reception, motivating dynamic resource-allocation frameworks. These frameworks jointly optimize user scheduling, transmission power, combined weights, and bandwidth \cite{peng2025dynamic}. Traffic offloading is another established strategy for the management of NTN-TN resources, explored in the literature through joint interference management and coordinated coexistence \cite{rahman2025joint} and traffic-aware scheduling with beam selection to increase throughput \cite{alam2024optimizing}, collectively improving energy efficiency, capacity, coverage, and operational flexibility \cite{Ozturk2026Sustainable}.

\textit{Interference Management for NTN--TN:} While the integration of NTN nodes into terrestrial cellular systems can improve spectrum utilization and extend coverage beyond terrestrial infrastructure, it also creates complex interference conditions due to the distinct propagation characteristics of terrestrial and non-terrestrial links \cite{Shang2026Spectrum}. Several approaches have been proposed to manage such interference. In \cite{rahman2023coexistence}, a stochastic-geometry-based framework optimizes exclusion-zone radii for terrestrial cellular and backhaul networks coexisting with fixed-satellite services at 28~GHz. Alternatively, interference from terrestrial BS sidelobes can be mitigated by steering spatial nulls towards visible satellites~\cite{kang2024terrestrial}. More recently, \cite{Allouch2026} optimized downlink beamforming to protect incumbent VSATs while maximizing terrestrial sum rate, showing that aggressive spatial nulling can unnecessarily sacrifice capacity; however, the formulation does not consider interference coordination across geographically separated BSs. In \cite{tsampazi2026satellite}, interference nulling and power control were considered for FR3 spectrum sharing, where terrestrial downlink transmissions interfere with LEO satellite uplinks. In \cite{jia2025joint}, a joint detection, channel-estimation, and interference-cancellation framework was developed for terrestrial--satellite downlink coexistence. Although these CSI-dependent approaches can reduce interference at NTN receivers, they require accurate interference-link CSI, increasing acquisition and processing overhead as the number of NTN receivers grows.

\textit{Digital Twin for NTNs:}
The dynamic nature of NTN environments, characterized by time-varying satellite visibility, beam coverage, and interference conditions, requires accurate real-time network representation. High-fidelity simulators, such as the one previously proposed in~\cite{Zhou2026UltraStar}, enable large-scale evaluation of LEO constellations but remain largely limited to offline analysis. DTN bridge this gap through real-time bidirectional interaction between physical and virtual networks, thus enabling network mapping and closed-loop decision making~\cite{wu2021digital,gao2023digital}. Recent advances include DT-assisted collaborative resource allocation for multi-satellite LEO systems through coordinated beam hopping and power allocation~\cite{Zhao2025BeamHopping}, as well as DT-supported satellite-terrestrial edge computing for improved resource sharing, reduced computation delay, and lower satellite energy consumption~\cite{ji2023cooperative}.

\textit{Uniqueness of the Proposed Work:}
Table~\ref{tab:related_works_summary} compares the present research with previous studies, most of which addressed only a subset of the resource-management aspects relevant to integrated NTN--TN systems. While some prior studies incorporate site-specific environmental information, its role is largely confined to channel modeling and performance evaluation rather than resource-control decisions. Motivated by these research gaps, we develop an O-RAN-enabled, DT-assisted interference-management framework for context-aware spectrum sharing in the FR2 band. Unlike prior studies, the proposed framework integrates multi-domain context-awareness, proactive interference management at satellite receivers through DT-based prediction and what-if analysis, and closed-loop resource coordination over standardized O-RAN interfaces. Collectively, these capabilities set apart the proposed framework from those previously proposed in the literature.

\begin{figure*}[!t]
    \centering
    \includegraphics[width=\textwidth]{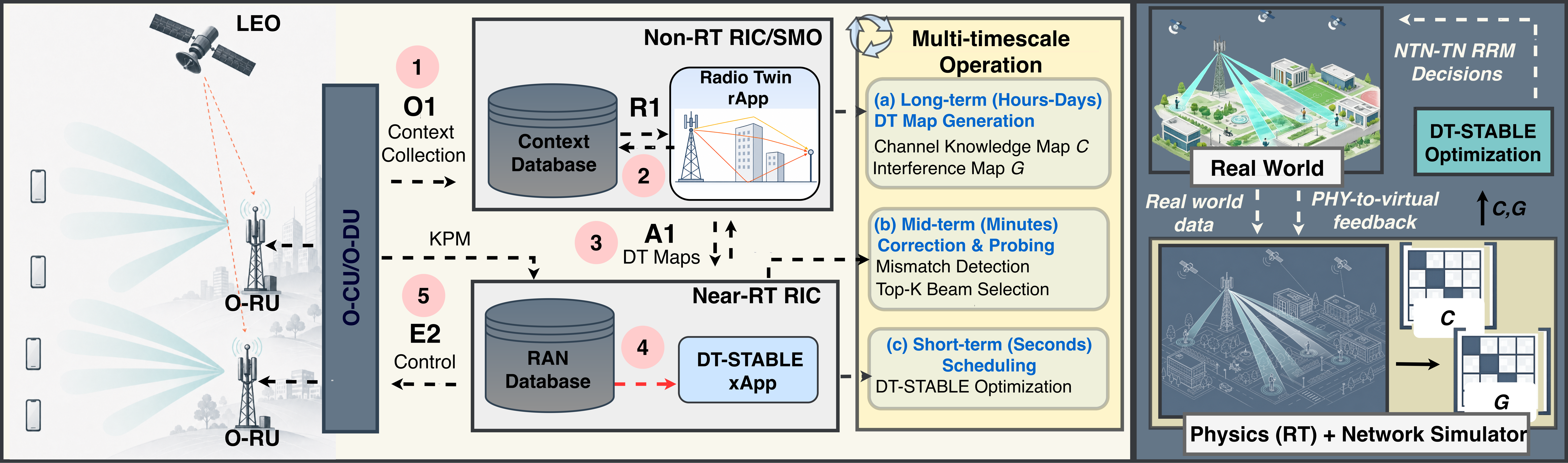}
    \caption{O-RAN-enabled DT-STABLE framework for multi-timescale context-aware NTN--TN coexistence.}
    \label{fig:oran_dt}
    \vspace{-0.4cm}
\end{figure*}

\vspace{-0.3cm}
\section{System Model}
\label{sec:system_model}

\subsection{System Overview}


We consider an NTN--TN coexistence scenario, where terrestrial downlink transmissions share frequency resources with an uplink LEO satellite system. The terrestrial segment follows the O-RAN architecture, comprising multiple open radio units (O-RUs) with antenna arrays, connected to open distributed unit (O-DU) and open central unit (O-CU) entities \cite{oran_architecture}. Radio resource control is supported by the near-real-time RIC (near-RT RIC), where xApps perform time-sensitive optimization, whereas the non-real-time RAN intelligent controller (non-RT RIC), hosted within the service management and orchestration (SMO) framework, provides non-real-time analytics, policy support, and DT-assisted contextual information. Let $\mathcal{J}=\{1,2,\ldots,J\}$ denote the set of O-RUs deployed over the considered geographic area. Furthermore, let $\mathcal{U}=\{1,2,\ldots,U\}$ denote the set of single-antenna user equipments (UEs). Each O-RU $j\in\mathcal{J}$ is equipped with a uniform rectangular array (URA) and a predefined discrete Fourier transform (DFT) beam codebook $\mathcal{B}_j=\{1,2,\ldots,K\}$ of candidate precoding vectors. Due to finite RF-chain and baseband processing resources, each O-RU can simultaneously activate only a limited number of beams \cite{molisch2017hybrid}. The satellite segment consists of a set $\mathcal{M}=\{1,2,\ldots,M\}$ of visible LEO satellites operating over the same spectrum. Each satellite $m\in\mathcal{M}$ is modeled as a victim receiver that must be protected from harmful aggregate interference generated by terrestrial downlink transmissions. Therefore, the coexistence scenario requires coordinated O-RU beam activation and UE association decisions that enhance terrestrial service quality while protecting the coexisting satellite segment from harmful interference.

In the proposed architecture, the CDT is maintained through the non-RT RIC using environmental, network, and satellite-context information. It provides two context-aware CKMs for near-RT control: a 3D geo-environment aware terrestrial coverage map storing the received power associated with each beam over predefined spatial cells, and an interference map that quantifies the coupling of each O-RU--beam pair with visible satellite receivers.
 During near-RT operation, each UE position is mapped to the corresponding coverage-map cell, and the stored received power is used to derive the achievable rate for candidate O-RU--beam pairs. Concurrently, the perceived interference from each O-RU-beam pair to visible satellites is obtained from the precomputed interference map. A purely measurement-driven approach would require $\mathcal{O}(UJK)$ and $\mathcal{O}(JKM)$ measurements per scheduling epoch. The proposed CDT instead reuses the precomputed maps and triggers only localized corrections when prediction mismatches occur, facilitating near-RT scalable operation.

\textbf{Assumptions:} We adopt the following practical assumptions. 
\textbf{A1:} UE position is available to the DT-assisted O-RAN control plane. This is consistent with emerging 6G positioning capabilities, where UE locations can be obtained through multi-modal sensing~\cite{alkhateeb2023deepsense} or GNSS and network-assisted fusion~\cite{brambilla2024integration}. This is so because future 6G systems are expected to jointly support communication, localization, and sensing. This information is used to query DT-generated coverage maps and support context-aware BS--beam selection. Sensitivity of DT-STABLE to localization errors is evaluated in Section~\ref{sec:performance_eval}. 
\textbf{A2:} The DT can accurately estimate the interference generated by each BS beam towards LEO satellite receivers. Residual errors may arise from propagation-model mismatch, environmental changes, or antenna-pattern uncertainty. This uncertainty is assumed to follow a lognormal distribution in the linear domain, equivalently represented by a zero-mean Gaussian perturbation in the dB domain. Accordingly, the actual interference is modeled as $I_{\mathrm{actual}}^{\mathrm{dB}} = I_{\mathrm{DT}}^{\mathrm{dB}} + \epsilon_I$, where $\epsilon_I \sim \mathcal{N}(0,\sigma_I^2)$ and where $\sigma_I$ \setlength{\skip\footins}{3pt}
denotes the standard deviation of the interference-prediction error in dB.%
\footnote{The Gaussian DT-error model is inspired by~\cite[Eq.~(12)]{bazzi2026dttraining}.}The impact of DT interference-prediction errors is also evaluated in Section~\ref{sec:performance_eval}.
\textbf{A3:} Inter-cell interference is neglected, which is reasonable for suburban macro deployments of downtilted BSs with typical inter-site distances and orthogonal frequency allocation among neighbouring cells via standard frequency reuse pattern \cite{rahman2025joint}. Intra-cell inter-beam interference among simultaneously active beams is also neglected, as motivated by the structural orthogonality of the DFT codebook. However, in practice, residual inter-beam interference can arise due to multipath-induced leakage, which will be considered in our future work.

\vspace{-0.35cm}
\subsection{DT-STABLE Framework}
\label{subsec:proposed_framework}

Fig.~\ref{fig:oran_dt} shows the architecture of the O-RAN-enabled, DT-assisted framework for interference-aware NTN--TN coexistence. Within the non-RT RIC, the radio-twin rApp constructs and maintains long-term DT maps using slowly varying contextual information. The DT-generated CKMs are periodically refined by identifying prediction mismatch with RAN measurements collected by the near-RT RIC. The DT-STABLE xApp then combines the updated CKMs with the near-RT RAN state to jointly optimize beam activation and UE association under interference constraints. This separation enables long-term map construction, mid-term feedback-driven refinement, and short-term resource optimization.

\textit{A Walk-Through of DT-STABLE:}
The operation of the proposed framework is shown in Fig.~\ref{fig:oran_dt}. First, network and environmental information are collected through the O1 interface and stored in the context database at the SMO/non-RT RIC layer \textbf{(Step~1)}. This information includes deployment maps, BS locations, beam codebooks, satellite ephemerides, weather information, and interference-management policies. Next, the radio twin rApp retrieves this information through the R1 interface and generates the channel knowledge map $C$ and the interference map $G$ \textbf{(Step~2)}. The generated DT maps are then delivered to the near-RT RIC through the A1 interface \textbf{(Step~3)}. At the near-RT RIC, the DT-STABLE xApp uses these maps together with RAN feedback received over the E2 interface, including UE positions, key performance measurements (KPMs), and measured rates. When a mismatch between DT-predicted and measured performance is detected, the xApp refines the DT information using limited real-world measurements before scheduling. Based on the updated DT and RAN information, the DT-STABLE xApp determines the beam activation and UE association decisions \textbf{(Step~4)}. These decisions are sent to the relevant E2 nodes through the E2 interface and applied to the physical RAN \textbf{(Step~5)}. The flow of \textbf{Steps~1--5} establishes a closed-loop and proactive operation that synchronizes the digital and physical networks across multiple timescales.
\vspace{-0.4cm}
\section{DT-Based Channel and Interference Modeling in NTN-TN Coexistence Networks}
\label{sec:dt_stable}

\subsection{Modeling of the Context-Aware Digital Twin (CDT)}
\label{subsec:cdt_modeling}

DT offers a practical approach for reducing dependence on extensive empirical measurements. This is done by synthesizing site-specific radio information that captures the underlying physical propagation environment~\cite{alkhateeb2023real,Luo2025DigitalTwinCSI}. In the present study, the CDT is modeled using the following components. \textbf{i) Real-world data:} The CDT relies on contextual information from the real-world to represent the site. In this article, this information includes site geometry, terrain and building data, BS/O-RU locations, UE locations, satellite visibility, antenna configurations, and beam codebooks. These inputs allow the CDT to characterize both terrestrial service links and interference paths towards the satellite uplink. \textbf{ii) Physics:} The physics layer models the site-specific TN and NTN propagation conditions relevant to coexistence control. Terrestrial O-RU coverage is generated using Sionna RT~\cite{hoydis2023sionnart} over a high-fidelity virtual 3D environment,\footnote{In this study, the virtual environment is constructed in Blender, incorporating BS/O-RU locations from OpenCell3D, building geometry from OpenStreetMap, and material properties assigned according to ITU standards.} spatial coverage cells, antenna configurations, and beamforming patterns. The resulting multipath components are combined to obtain channel responses associated with the considered beams over the coverage region. During near-RT operation, UE locations are used only to identify the corresponding coverage-map cells. For the NTN links, O-RU--satellite RF signal propagation is modeled using OpenNTN~\cite{Due2025OpenNTN}. Satellite positions are obtained from TLE data, while link losses are determined from the antenna responses, elevation angle, and atmospheric attenuation. \textbf{iii) Network simulator:}
Instead of emulating the full network stack, the proposed CDT adopts a goal-specific link-level abstraction for NTN--TN coexistence control. The terrestrial propagation outputs are used to obtain the received-power information over the spatial coverage cells, while the NTN propagation outputs determine interference coupling between each O-RU beam and visible satellite receiver. These quantities are stored in terrestrial coverage map $\mathcal{C}$ and satellite interference map $\mathcal{G}$, respectively, and are then used to compute the performance metrics, such as downlink throughput and aggregated interference. \textbf{\mbox{iv) Physical-to-virtual feedback:}} The physical-to-virtual feedback part transfers selected RAN observations back to the CDT. In the proposed O-RAN architecture, the near-RT RIC obtains UE locations, KPM reports, and measured link rates through the E2 interface. These observations are then compared to the CDT-predicted rates computed from the current channel map $C$. When the predictions deviate from the physical RAN observations, relevant link entries are refined using the latest feedback. This creates a mid-timescale synchronization loop between physical RAN and the CDT, avoiding full map regeneration at every scheduling interval. \textbf{\mbox{v) DT-assisted resource control:}}  The DT-STABLE xApp uses the current CDT state together with near-RT RAN feedback to determine beam activation and UE association decisions under the satellite interference constraints. The resulting control actions are then delivered to the physical RAN through the O-RAN control loop. 

\vspace{-0.3cm}
\subsection{TN Antenna Modelling for Coexistence Analysis}
\label{subsec:tn_model}

Each terrestrial BS is equipped with an $N_x\times N_y$ URA and uses a finite DFT codebook for directional transmission. The antenna elements are separated by half-wavelength spacing, i.e., $d_x=d_y=\lambda/2$. Each antenna element follows the 3GPP TR~38.901 element pattern \cite{3gpp_tr38901}. For a direction specified by elevation
angle $\theta$ and azimuth angle $\phi$, the element gain is modeled as shown below:

{\footnotesize
\begin{equation}
G_{\mathrm{elem}}^{3\mathrm{GPP}}(\theta,\phi)
=
G_{\max}
-
\min\left\{
12\left(\frac{\phi}{\phi_{3\mathrm{dB}}}\right)^2
+
12\left(\frac{\theta-90^\circ}{\theta_{3\mathrm{dB}}}\right)^2,
A_m
\right\}
\label{eq:3gpp_element_gain}
\end{equation}
}
where $G_{\max}$ is the maximum element gain, $\phi_{3{\rm dB}}$ and
$\theta_{3{\rm dB}}$ are the horizontal while vertical half-power beamwidths,
and $A_m$ is the maximum attenuation.

The $k$-th entry of BS $j$ is represented by a unit-norm DFT precoding vector
$\mathbf{f}_{j,k}\in\mathbb{C}^{N_xN_y\times 1}$. Let
$\mathbf{a}_{j}(\theta,\phi)\in\mathbb{C}^{ML\times 1}$ denote the URA steering response in direction $(\theta,\phi)$, including the antenna element
locations and BS orientation. The array gain induced by the $k$-th precoding vector is
given by the following:
\begin{equation}
G_{\rm array}^{j,k}(\theta,\phi)
=
10\log_{10}
\left(
\left|
\mathbf{a}_{j}^{\mathsf{H}}(\theta,\phi)
\mathbf{f}_{j,k}
\right|^2
\right).
\label{eq:array_precoding_gain}
\end{equation}

Accordingly, the directional gain of the $k$-th terrestrial beam of BS $j$ is computed as follows:
\begin{equation}
G^{\rm TN}_{j,k}(\theta,\phi)
=
G_{\rm elem}^{\rm 3GPP}(\theta,\phi)
+
G_{\rm array}^{j,k}(\theta,\phi).
\label{eq:tn_directional_gain}
\end{equation}
For the BS--satellite direction $(\theta_{j,m}(t),\phi_{j,m}(t))$, the
corresponding beam-dependent gain is computed as shown below:
\begin{equation}
G^{\rm TN}_{j,k,m}(t)
=
G^{\rm TN}_{j,k}
\big(\theta_{j,m}(t),\phi_{j,m}(t)\big).
\label{eq:tn_gain_sat_direction}
\end{equation}

\vspace{-0.3cm}
\subsection{NTN Modelling for Coexistence Analysis}
\label{subsec:ntn_model}

\subsubsection{NTN Antenna Modelling}
\label{subsubsec:ntn_antenna_model}

The receive antenna of each LEO satellite is modeled using a
circular-aperture reflector pattern~\cite{3gpp_tr38811}. Let
$\psi_{j,m}(t)$ denote the off-axis angle between the boresight
of the receive antenna of satellite $m$ and the direction of
the signal arriving from BS $j$. The corresponding normalized
antenna power pattern is given by Eq. (\ref{eq:ntn_antenna_pattern}):

\begingroup
\footnotesize
\begin{equation}
F_m\!\left(\psi_{j,m}(t)\right)
=
\begin{cases}
1,
& \psi_{j,m}(t)=0,\\[1mm]
4\left|
\dfrac{
J_1\!\left(k_0a_m\sin\!\left(\psi_{j,m}(t)\right)\right)}
{k_0a_m\sin\!\left(\psi_{j,m}(t)\right)}
\right|^2,
& 0<\left|\psi_{j,m}(t)\right|\leq 90^\circ,\\[3mm]
0,
& \text{otherwise},
\end{cases}
\label{eq:ntn_antenna_pattern}
\end{equation}
\endgroup
where $J_1(\cdot)$ is the first-order Bessel function of the
first kind, $a_m$ is the antenna-aperture radius of satellite
$m$, while $k_0=2\pi f_c/c$ is the wave number, with $f_c$ and
$c$ denoting the carrier frequency and the speed of light,
respectively. The pattern is normalized such that
$F_m(0)=1$ along the antenna boresight.
Assuming that the receiver noise figure and antenna noise
temperature are independent of the signal arrival direction,
the directional $G/T$ of satellite $m$ towards BS $j$ is modeled as follows:
\begin{equation}
\small
\left(\frac{G}{T}\right)_{j,m}(t)
=
\left(\frac{G}{T}\right)_{m,\max}
+
10\log_{10}
F_m\!\left(\psi_{j,m}(t)\right),
\label{eq:directional_satellite_gt}
\end{equation}
where $\left(G/T\right)_{m,\max}$ denotes the maximum
boresight gain-to-noise-temperature ratio of satellite $m$.

\begin{figure*}[!t]
    \centering
    \includegraphics[width=0.87\textwidth]{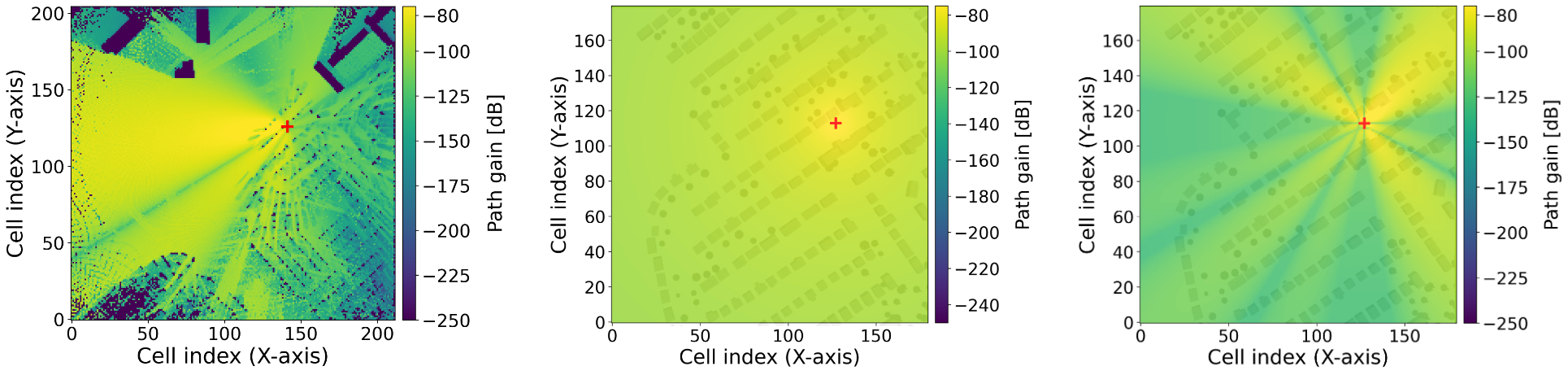}
    \caption{Comparison of CKM construction approaches: (a) environment-aware multi-beam CKM; (b) environment-unaware CKM (with omnidirectional beam); and (c) environment-unaware multi-beam CKM.}
    \label{fig:construction_ckm}
    \vspace{-0.7cm}
\end{figure*}
\subsubsection{NTN Channel Model}
\label{subsec:ntn_channel}

The NTN propagation channel is generated offline using OpenNTN
\cite{Due2025OpenNTN}, an open-source implementation of the 3GPP
TR~38.811 geometry-based stochastic channel model (GSCM)
\cite{3gpp_tr38811}. For each BS--satellite link and ephemeris
snapshot, channel generation is conditioned on the corresponding
link geometry and deployment scenario. The GSCM determines the
LOS/NLOS propagation state and generates the associated large-scale
channel parameters, including shadow fading, delay and angular spreads,
and the Rician $K$-factor. Conditioned on these parameters, the
small-scale channel realization is synthesized from multipath clusters
comprising rays with prescribed delays, powers, and angles of arrival
and departure. The temporal evolution associated with satellite
mobility is incorporated through the time-varying link geometry and
the resulting Doppler characteristics.

For each BS--satellite pair $(j,m)$, the link geometry is determined
from the BS position $\mathbf{q}_j\in\mathbb{R}^3$ and the
time-varying satellite position $\mathbf{q}_m(t)\in\mathbb{R}^3$,
expressed in a common three-dimensional coordinate system. The
satellite position is obtained from its ephemeris data, and the
corresponding BS--satellite slant distance is
$d_{j,m}(t)=\|\mathbf{q}_m(t)-\mathbf{q}_j\|_2$.

The BS--satellite geometry is further characterized by the elevation
angle with respect to the local vertical at the BS. Let
$\mathbf{n}_j$ denote the local vertical unit vector at BS $j$. The
elevation angle of satellite $m$ observed from BS $j$ is given by
\begin{equation}
\theta_{j,m}(t)
=
\sin^{-1}\left(
\frac{
\left(\mathbf{q}_m(t)-\mathbf{q}_j\right)^{\mathrm T}
\mathbf{n}_j
}{
d_{j,m}(t)
}
\right).
\label{eq:satellite_elevation}
\end{equation}

For directional antenna modeling, the BS--satellite direction is
expressed in the local coordinate frame of the BS antenna panel,
yielding the zenith and azimuth angles
$\vartheta_{j,m}^{\mathrm{loc}}(t)$ and
$\phi_{j,m}^{\mathrm{loc}}(t)$, respectively. These angles account for
the panel orientation and downtilt. Let
$\mathbf{f}_{j,k}\in\mathbb{C}^{N_{\mathrm{ant}}\times 1}$ denote the
unit-norm precoding vector of beam $k$ at BS $j$, and let
$\mathbf{a}_{j}(\vartheta,\phi)$ denote the corresponding array
response, normalized such that
$\|\mathbf{a}_{j}(\vartheta,\phi)\|_{2}^{2}=N_{\mathrm{ant}}$.
The directional transmit gain of beam $k$ toward satellite $m$ is
then given by
\begin{equation}
\begin{aligned}
G_{j,k,m}^{\mathrm{tx}}(t)
&=
G_{\mathrm{E},j}\!\left(
\vartheta_{j,m}^{\mathrm{loc}}(t),
\phi_{j,m}^{\mathrm{loc}}(t)
\right)\\
&\quad+
10\log_{10}\!\left|
\mathbf{a}_{j}^{\mathsf{H}}\!\left(
\vartheta_{j,m}^{\mathrm{loc}}(t),
\phi_{j,m}^{\mathrm{loc}}(t)
\right)
\mathbf{f}_{j,k}
\right|^{2},
\end{aligned}
\label{eq:ntn_directional_tx_gain}
\end{equation}
where $G_{\mathrm{E},j}(\vartheta,\phi)$ denotes the antenna-element
gain in dBi. Thus, \eqref{eq:ntn_directional_tx_gain} accounts for
both the element radiation pattern and the beam-dependent array factor
in the satellite direction.

The large-scale propagation loss between BS $j$ and satellite $m$ is
modeled as
\begin{equation}
PL_{j,m}^{\mathrm{NTN}}(t)
=
PL_{j,m}^{\mathrm{b}}(t)
+
PL_{j,m}^{\mathrm{g}}(t)
+
PL_{j,m}^{\mathrm{s}}(t),
\label{eq:ntn_total_pathloss}
\end{equation}
where $PL_{j,m}^{\mathrm{b}}(t)$ denotes the basic path loss,
$PL_{j,m}^{\mathrm{g}}(t)$ denotes atmospheric attenuation, and
$PL_{j,m}^{\mathrm{s}}(t)$ denotes the applicable scintillation loss.
Building-entry loss is omitted since the terrestrial BSs are deployed
outdoors.

The basic path loss is expressed as
\begin{equation}
\begin{aligned}
PL_{j,m}^{\mathrm{b}}(t)
&=
32.45
+20\log_{10}\!\left(f_{\mathrm{c}}[\mathrm{MHz}]\right)
+20\log_{10}\!\left(d_{j,m}(t)[\mathrm{km}]\right)\\
&\quad
+SF_{j,m}(t)
+CL_{j,m}(t),
\end{aligned}
\label{eq:ntn_basic_pathloss}
\end{equation}
where the first three terms correspond to the free-space path loss.
Let
$\chi_{j,m}(t)\in\{\mathrm{LOS},\mathrm{NLOS}\}$
denote the propagation condition of the BS--satellite link, determined
according to the scenario- and elevation-dependent LOS probability.
Conditioned on $\chi_{j,m}(t)$, the shadow-fading term is modeled as
\begin{equation}
SF_{j,m}(t)
\sim
\mathcal{N}\!\left(
0,
\sigma_{\mathrm{SF}}^{2}
\!\left(
\mathcal{E},
\theta_{j,m}(t),
\chi_{j,m}(t)
\right)
\right),
\end{equation}
where $\mathcal{E}$ denotes the deployment scenario. The corresponding
clutter loss is
\begin{equation}
CL_{j,m}(t)=
\begin{cases}
0, & \chi_{j,m}(t)=\mathrm{LOS},\\[1mm]
CL_{\mathcal{E}}\!\left(\theta_{j,m}(t),f_{\mathrm{c}}\right),
& \chi_{j,m}(t)=\mathrm{NLOS},
\end{cases}
\label{eq:ntn_clutter_loss}
\end{equation}
where $CL_{\mathcal{E}}(\cdot)$ denotes the scenario-dependent NLOS
clutter loss. The atmospheric component
$PL_{j,m}^{\mathrm{g}}(t)$ accounts for gaseous, rain, and cloud
attenuation along the Earth--space path according to the adopted
3GPP and ITU-R propagation models \cite{3gpp_tr38811,ITU-R-P618-14}.
For the considered configuration, the applicable scintillation term is
set according to the corresponding 3GPP/ITU-R model.

For each BS--satellite geometry, the GSCM further generates
$N_{\mathrm{MC}}$ independent stochastic channel realizations,
with the link geometry held fixed. These realizations account for the
small-scale multipath structure through the corresponding cluster and
ray parameters, including delays, powers, angular characteristics,
cross-polarization ratios, and random phases. Let
$\ell_{j,k,m}^{(r)}(t)$ denote the effective power coupling between
beam $k$ of BS $j$ and satellite $m$ obtained from the $r$th
realization. The NTN coupling retained for interference evaluation is
defined as follows:
\vspace{-0.3cm}
\begin{equation}
\ell_{j,k,m}^{\mathrm{NTN}}(t)
=
\frac{1}{N_{\mathrm{MC}}}
\sum_{r=1}^{N_{\mathrm{MC}}}
\ell_{j,k,m}^{(r)}(t),
\qquad
N_{\mathrm{MC}}=50.
\label{eq:ntn_channel_gain}
\vspace{-0.3cm}
\end{equation}

\vspace{-0.3cm}
\subsection{Construction of Multi-Antenna CKM}
\label{subsec:ckm}

Let $\mathcal{A}$ denote the considered coverage region, which is partitioned into $N_{\mathrm{c}}$ non-overlapping rectangular cells $\{Q_n\}_{n=1}^{N_{\mathrm{c}}}$ of size $\Delta_x \times \Delta_y$, where $\Delta_x=\Delta_y=5$ m in the considered implementation. Let
$\mathcal{N}_{\mathrm{c}}=\{1,\ldots,N_{\mathrm{c}}\}$ denote the
corresponding cell-index set. Let
$\mathcal{B}_j=\{1,\ldots,K_j\}$ denote the beam-index set of
O-RU/gNB site $j$, where
$\mathbf{f}_{j,k}\in\mathbb{C}^{N_{\mathrm{ant}}\times 1}$ is the
precoding vector associated with beam $k\in\mathcal{B}_j$. Each O-RU/gNB site is equipped with an $N_x\times N_y$ URA, comprising $N_{\mathrm{ant}}=N_xN_y$ antenna elements, and uses a finite DFT beamforming codebook. The $k$-th precoding vector at site $j$ is expressed as
$\mathbf{f}_{j,k}=
(\mathbf{a}_x(p)\otimes\mathbf{a}_y(q))/\sqrt{N_{\mathrm{ant}}}$,
where $\mathbf{a}_x(p)$ and $\mathbf{a}_y(q)$ denote the horizontal
and vertical DFT steering vectors, respectively; furthermore, $(p,q)$ is the corresponding two-dimensional DFT index pair, while $\otimes$ denotes
the Kronecker product. For the full DFT codebook,
$K_j=N_xN_y=N_{\mathrm{ant}}$.

We perform extensive site-specific high-fidelity RT to generate an offline coverage map for every candidate beam at each O-RU/gNB site. For cell $n$, site $j$, and beam $k$, the average beamformed path gain is given by the following:
\vspace{-0.3cm}
\begin{equation}
g_{n,j,k}
=
\frac{1}{|\mathcal{Q}_n|}
\int_{\mathcal{Q}_n}
\left|
\mathbf{h}_{j}^{H}(\mathbf{s})\mathbf{f}_{j,k}
\right|^{2}
\,\mathrm{d}\mathbf{s},
\label{eq:ckm_cell_path_gain}
\end{equation}
where $|\mathcal{Q}_n|$ denotes the area of cell $\mathcal{Q}_n$,
$\mathbf{s}=(x,y)\in\mathcal{Q}_n$ denotes a spatial location within
the cell, $\mathrm{d}\mathbf{s}=\mathrm{d}x\,\mathrm{d}y$ is the
infinitesimal surface element, and $\mathbf{h}_{j}(\mathbf{s})$
denotes the channel vector between site $j$ and, finally, location
$\mathbf{s}$. The resulting CKM enables DT-STABLE to proactively
evaluate terrestrial service-side utilities, such as the received signal-to-noise ratio (SNR) and
achievable spectral efficiency, for each candidate
O-RU--beam--UE association.

In this study, the complete multi-antenna CKM is represented by the
beam-dependent received signal strength (RSS) tensor
$\mathbf{C}=\{\rho_{n,j,k}\mid n\in\mathcal{N}_{\mathrm{c}},\,
j\in\mathcal{J},\,k\in\mathcal{B}_j\}$.
where
$\rho_{n,j,k}=P_j^{b}g_{n,j,k}$ denotes the RSS in the linear domain
for beam $k$ of site $j$ over cell $n$, while $P_j^{b}$ is the fixed
transmit power allocated to one beam.
During near-RT operation, the UE position is mapped to the corresponding coverage-map cell, and the stored RSS associated with each beam is used to approximate the received power at the UE location. The accuracy of this approximation depends on the cell size and the spatial variation of the received power within the cell.

Fig.~\ref{fig:construction_ckm} compares three CKM models with varying levels
of environmental and beam awareness. The environment-aware multi-beam CKM captures site-specific propagation conditions for each spatial cell and candidate beam. By contrast, the
environment-unaware CKM relies only on location, and its multi-beam variant includes beam dependence without real propagation effects, thus generating idealistic uniform beam coverage. By incorporating site-specific information (blockage, terrain, and multi-path propagation), environment-aware multi-beam CKM can provide spatially consistent, physics-accurate channel gain at any UE position within the O-RU coverage for each beam.
\vspace{-0.44cm}
\subsection{Interference Map Construction}
\label{subsec:interference_modeling}

We model the aggregate interference generated by active terrestrial BS beams
toward each visible LEO satellite receiver. Let $P_j^{b}$ denote the fixed
transmit power allocated to each active beam of BS $j$. Combining the effective
NTN power coupling in Eq.~\eqref{eq:ntn_channel_gain} with the directional
satellite gain-to-noise-temperature ratio in
Eq.~\eqref{eq:directional_satellite_gt}, the INR contribution of beam $k$ at
BS $j$ toward satellite $m$ is expressed in decibels as
\vspace{-0.1cm}
\begin{equation}
\small
\begin{aligned}
I_{j,k,m}^{\mathrm{dB}}(t)
&=
P_j^{b,\mathrm{dBW}}
+10\log_{10}\!\left(\ell_{j,k,m}^{\mathrm{NTN}}(t)\right) \\
&\quad
+\bigl(G/T\bigr)_{j,m}(t)
-10\log_{10}(\kappa B).
\end{aligned}
\label{eq:per_beam_inr}
\end{equation}
where $P_j^{b,\mathrm{dBW}}$ is the per-beam transmit power in dBW, $\bigl(G/T\bigr)_{j,m}(t)$ is expressed in dB/K, $B$ is the bandwidth in Hz, and $\kappa$ is the Boltzmann constant. Accordingly, linear INR coupling coefficient is defined as follows:
\vspace{-0.4cm}

{\small
\begin{equation}
g_{j,k,m}(t)
=10^{I_{j,k,m}^{\mathrm{dB}}(t)/10}
=\frac{P_j^{b}\ell_{j,k,m}^{\mathrm{NTN}}(t)}{\kappa B}
10^{\left(G/T\right)_{j,m}(t)/10},
\label{eq:linear_inr_coupling}
\end{equation}
}
where $P_j^{b}$ is expressed in watts and $\ell_{j,k,m}^{\mathrm{NTN}}(t)$ is the effective NTN power coupling defined in Eq. \eqref{eq:ntn_channel_gain}. Thus, $g_{j,k,m}(t)$ represents the linear INR contribution of beam $k$ of BS $j$ toward satellite $m$. 

Next, let $s_{j,k}(t)\in\{0,1\}$ indicate whether beam $k$ of BS $j$ is active. Assuming that all active beams overlap the considered bandwidth, transmit with unit activity, and contribute noncoherently at the satellite receiver, the aggregate INR at satellite $m$ is obtained by summing the individual contributions in the linear domain as follows:
\begin{equation}
I_m(t)=\sum_{j\in\mathcal{J}}\sum_{k\in\mathcal{B}_j}s_{j,k}(t)g_{j,k,m}(t).
\label{eq:aggregate_inr}
\vspace{-0.2cm}
\end{equation}

\begin{figure}[t]
    \centering

    \includegraphics[
        width=0.86\columnwidth,
        trim=0 2mm 0 0,
        clip
    ]{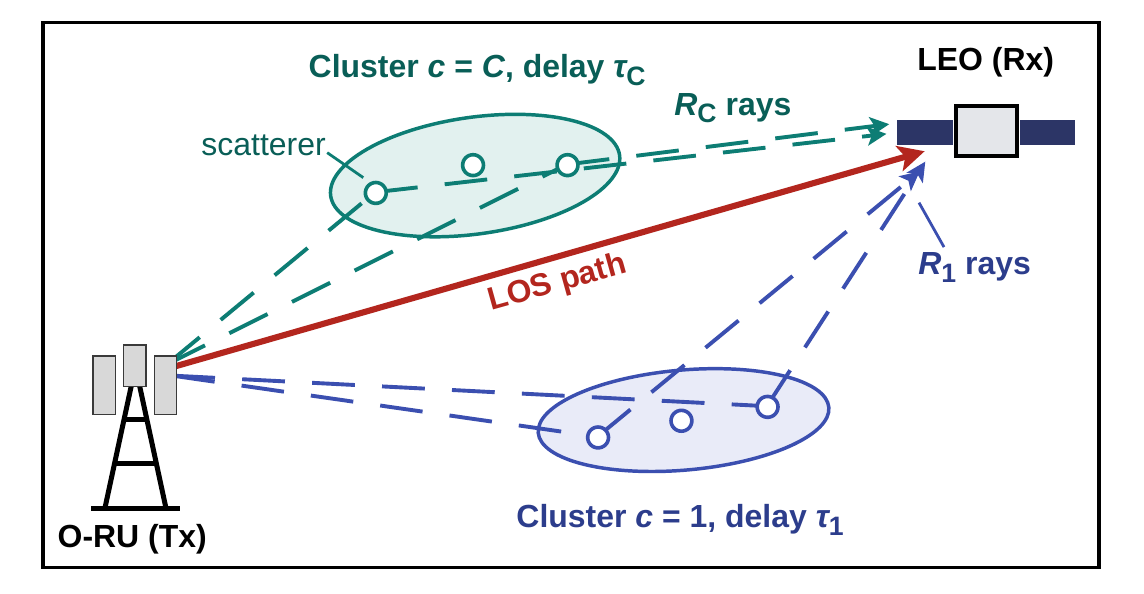}
    \par\vspace{-1.5mm}
    {\footnotesize (a)}
    \par\vspace{-1mm}

    \includegraphics[
        width=0.86\columnwidth,
        trim=0 2mm 0 0,
        clip
    ]{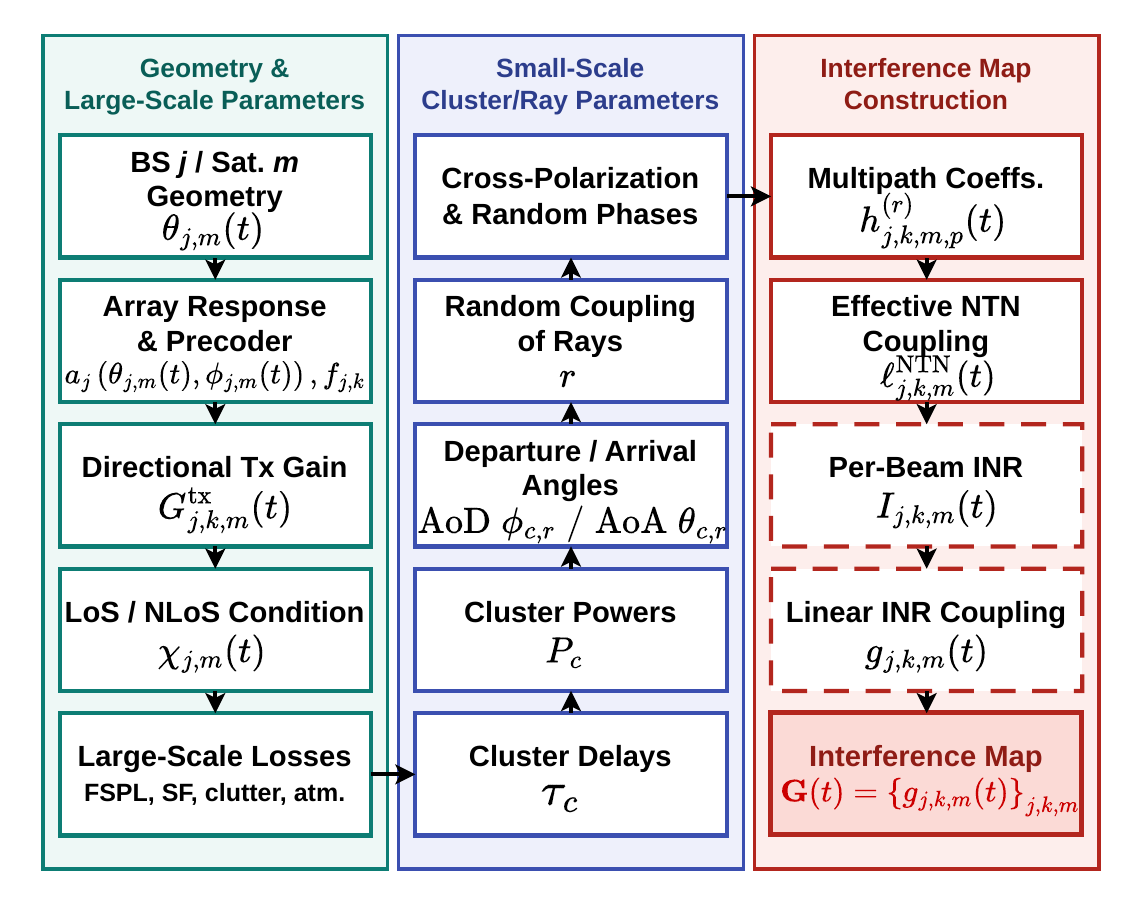}
    \par\vspace{-1.5mm}
    {\footnotesize (b)}
    \par\vspace{-0.5mm}

    \caption{NTN interference-map construction:
    (a) BS-to-LEO uplink channel model;
    (b) geometry, channel generation, and interference-map pipeline.}
    \label{fig:inr_pipeline}
    \vspace{-0.1cm}
\end{figure}

The interference map provides a predictive representation of the interference coupling between terrestrial BS beams and visible satellite receivers over the satellite trajectory. By combining TLE-based satellite geometry with the directional antenna responses and NTN propagation model, the map captures the time-varying interference exposure associated with each candidate BS--beam pair. Fig.~\ref{fig:inr_pipeline} illustrates the corresponding BS-to-LEO channel representation and the overall geometry-, channel-, and interference-map construction pipeline. This information enables DT-STABLE to anticipate interference events that may violate the satellite protection threshold and account for satellite protection in the near-RT resource-allocation policy.
\vspace{-0.2cm}
\section{Radio Resource Management for Harmonious NTN--TN Coexistence}

\subsection{Optimization Problem Formulation}
\label{sec:optimization}

In this section, we formulate the joint beam activation and UE association problem underpinning the proposed NTN–TN coexistence framework. The formulation balances terrestrial service performance and satellite protection, while promoting efficient use of available beam resources.

\textbf{Decision Variables:} We define binary beam activation variable 
$s_{j,b}\in\{0,1\}$, for all $j\in\mathcal{J}$ and $b\in\mathcal{B}_j$, 
where $s_{j,b}=1$ indicates that beam $b$ of BS $j$ is active. 
We also define binary association variable $x_{u,j,b}\in\{0,1\}$, 
for all $u\in\mathcal{U}$, $j\in\mathcal{J}$, and $b\in\mathcal{B}_j$, 
where $x_{u,j,b}=1$ indicates that UE $u$ is served by BS $j$ through beam $b$. The load of beam $b$ at BS $j$ is defined as
$K_{j,b}=\sum_{u\in\mathcal{U}}x_{u,j,b}$, for all
$j\in\mathcal{J}$ and $b\in\mathcal{B}_j$. Next, we define the
achievable downlink rate for each candidate UE--BS--beam association.
Let $P_j^{\mathrm{b}}$ denote the transmit power allocated to one
active beam of BS $j$. Let $n_u$ denote the coverage-map cell associated with UE $u$. The received power for candidate BS--beam pair $(j,k)$ is obtained from the CKM as $\rho_{n_u,j,k}$. Accordingly, the received SNR is given by $\gamma_{u,j,k}=\rho_{n_u,j,k}/\sigma_u^2$, where $\sigma_u^2=N_0BF_{\mathrm{N}}$ is the receiver noise power, $N_0$ is the thermal noise power spectral density, $B$ is the system bandwidth, and, finally, $F_{\mathrm{N}}$ is the receiver noise factor. The corresponding achievable spectral efficiency is $c_{u,j,k}=\log_2(1+\gamma_{u,j,k})$ bps/Hz. To capture system throughput and user-level fairness, we adopt a logarithmic utility formulation~\cite{ye2013user}. We first define the load-aware terrestrial network utility as follows:
\begin{equation}
\label{eq:utility}
U(\mathbf{x}) \triangleq
\sum_{u\in\mathcal{U}}
\sum_{j\in\mathcal{J}}
\sum_{b\in\mathcal{B}_j}
x_{u,j,b}
\log\!\left(
\frac{c_{u,j,b}}{K_{j,b}}
\right).
\end{equation}

To capture the trade-off between terrestrial network utility and
satellite interference protection, the joint optimization problem is
formulated as shown in Eq. \eqref{eq:P0}:
\begin{equation}
\label{eq:P0}
\begin{aligned}
\textbf{P0:}\quad
\max_{\mathbf{x},\mathbf{s}}\quad
&
U(\mathbf{x})
-
\sum_{m\in\mathcal{M}}
\lambda_{\max,m}
\left[
I_m-I_m^{\mathrm{th}}
\right]_{+}
\\[-0.2em]
\mathrm{s.t.}\quad
&
\left\{
\begin{aligned}
\textbf{C1:}\;&
\sum_{j\in\mathcal{J}}\sum_{b\in\mathcal{B}_j}
x_{u,j,b}=1,
&& \forall u\in\mathcal{U},
\\
\textbf{C2:}\;&
x_{u,j,b}\leq s_{j,b},
&& \forall u,j,b,
\\
\textbf{C3:}\;&
\sum_{b\in\mathcal{B}_j}s_{j,b}\leq N_{\mathrm{RF}},
&& \forall j\in\mathcal{J},
\\
\textbf{C4:}\;&
K_{j,b}=\sum_{u\in\mathcal{U}}x_{u,j,b},
&& \forall j,b,
\\
\textbf{C5:}\;&
x_{u,j,b},\,s_{j,b}\in\{0,1\},
&& \forall u,j,b.
\end{aligned}
\right.
\end{aligned}
\end{equation}

where $[z]_{+}\triangleq \max(z,0)$ and $I_m$ denotes the aggregate
linear-domain INR at satellite $m$, as defined in
\eqref{eq:aggregate_inr}. The coefficient $\lambda_{\max,m}\geq0$ controls the
trade-off between terrestrial network utility and interference
protection for satellite $m$. Therefore, no interference penalty is
incurred when $I_m\leq I_m^{\mathrm{th}}$, whereas violations above the
threshold are increasingly penalized as $\lambda_{\max,m}$ increases.

\begin{algorithm}[!t]
\caption{DT-STABLE Optimization}
\label{alg:dt_stable}
\vspace{-1mm}
\tiny
\DontPrintSemicolon
\SetAlgoLined
\scriptsize
\SetAlFnt{\scriptsize}
\SetAlCapFnt{\scriptsize}
\SetAlCapNameFnt{\scriptsize}
\SetAlgoNlRelativeSize{-2}
\setlength{\algomargin}{0.1em}
\SetInd{0.05em}{0.15em}
\setlength{\interspacetitleruled}{1.5pt}
\setlength{\interspacealgoruled}{1.5pt}
\setlength{\algotitleheightrule}{0.2pt}
\setlength{\algoheightrule}{0.2pt}
\setlength{\textfloatsep}{2pt}
\setlength{\floatsep}{2pt}
\setlength{\intextsep}{2pt}
\setlength{\abovedisplayskip}{1pt}
\setlength{\belowdisplayskip}{1pt}
\setlength{\abovedisplayshortskip}{1pt}
\setlength{\belowdisplayshortskip}{1pt}

\KwIn{
Candidate beam sets $\{\mathcal{F}_j\}_{j\in\mathcal{J}}$;
DT-predicted rates $\{c_{u,j,b}\}$;
DT-predicted interference couplings $\{g_{j,b,m}\}$;
interference thresholds $\{I_m^{\rm th}\}$;
interference-price caps $\{\lambda_{\max,m}\}$;
RF-chain limit $N_{\rm RF}$;
step-size parameters $\alpha_0,\beta,c,\gamma,\alpha_{\max}$;
tolerance $\varepsilon$;
maximum number of iterations $T_{\max}$.
}

\KwOut{
Beam activation $\{s_{j,b}^{\star}\}$ and user association
$\{x_{u,j,b}^{\star}\}$.
}

\textbf{Initialization:}
Set $\lambda_m^{(0)}=0$, $\forall m\in\mathcal{M}$, and
$\mu_{j,b}^{(0)}=0$, $\forall j\in\mathcal{J}$,
$b\in\mathcal{F}_j$;
set $t=0$ and $\Delta=\infty$.\;

\While{$\Delta>\varepsilon$ and $t<T_{\max}$}{

Set $x_{u,j,b}^{(t)}=0$, $\forall u\in\mathcal{U}$,
$j\in\mathcal{J}$, $b\in\mathcal{F}_j$.\;

\ForEach{$u\in\mathcal{U}$}{
Select the tentative BS--beam pair
\[
(j_u^\star,b_u^\star)
\in
\arg\max_{\substack{j\in\mathcal{J}\\ b\in\mathcal{F}_j}}
\left\{
\log(c_{u,j,b})
-\mu_{j,b}^{(t)}
\right\}.
\]
Set $x_{u,j_u^\star,b_u^\star}^{(t)}=1$.\;
}

Set $s_{j,b}^{(t)}=0$, $\forall j\in\mathcal{J}$,
$b\in\mathcal{F}_j$.\;

\ForEach{$j\in\mathcal{J}$}{

\ForEach{$b\in\mathcal{F}_j$}{
Compute the price-adjusted beam score
\[
Q_{j,b}^{(t)}
=
\sum_{u\in\mathcal{U}}
x_{u,j,b}^{(t)}
\left[
\log(c_{u,j,b})
-\mu_{j,b}^{(t)}
\right]
-
\sum_{m\in\mathcal{M}}
\lambda_m^{(t)}g_{j,b,m}.
\]
}

Select $\mathcal{A}_j^{(t)}$ as the set of at most
$N_{\rm RF}$ beams in $\mathcal{F}_j$ having the largest
$Q_{j,b}^{(t)}$ values among those satisfying
\[
\sum_{u\in\mathcal{U}}x_{u,j,b}^{(t)}>0.
\]

Set $s_{j,b}^{(t)}=1$, $\forall b\in\mathcal{A}_j^{(t)}$.\;

}

Define the globally active BS--beam set
\[
\mathcal{A}^{(t)}
=
\left\{
(j,b):
j\in\mathcal{J},\;
b\in\mathcal{A}_j^{(t)}
\right\}.
\]

\ForEach{$u\in\mathcal{U}$ assigned to some
$(j,b)\notin\mathcal{A}^{(t)}$}{

Set $x_{u,j,b}^{(t)}=0$ for the discarded assignment.\;

Select the globally active BS--beam pair
\[
(j_u',b_u')
\in
\arg\max_{(j,b)\in\mathcal{A}^{(t)}}
\left\{
\log(c_{u,j,b})
-\mu_{j,b}^{(t)}
\right\}.
\]

Set $x_{u,j_u',b_u'}^{(t)}=1$.\;

}

\ForEach{$m\in\mathcal{M}$}{

Compute the aggregate interference and residual
\[
I_m^{(t)}
=
\sum_{j\in\mathcal{J}}
\sum_{b\in\mathcal{F}_j}
s_{j,b}^{(t)}g_{j,b,m},
\qquad
v_m^{(t)}
=
I_m^{(t)}-I_m^{\rm th}.
\]

Compute the adaptive step size
\[
\alpha_m^{(t)}
=
\min
\left\{
\frac{\alpha_0}{t+1}
\left[
1+c
\left(
e^{\gamma\max\{0,v_m^{(t)}\}}-1
\right)
\right],
\alpha_{\max}
\right\}.
\]

Update the interference price
\[
\lambda_m^{(t+1)}
=
\min
\left\{
\lambda_{\max,m},
\max
\left\{
0,
\lambda_m^{(t)}
+
\alpha_m^{(t)}v_m^{(t)}
\right\}
\right\}.
\]

}

\ForEach{$j\in\mathcal{J}$}{
\ForEach{$b\in\mathcal{F}_j$}{

Compute the beam demand and price-induced supply
\[
K_{j,b}^{\rm dem,(t)}
=
\sum_{u\in\mathcal{U}}
x_{u,j,b}^{(t)},
\qquad
K_{j,b}^{\rm sup,(t)}
=
\min
\left\{
|\mathcal{U}|,
\exp(\mu_{j,b}^{(t)}-1)
\right\}.
\]

Update the load price
\[
\mu_{j,b}^{(t+1)}
=
\mu_{j,b}^{(t)}
-
\frac{\beta}{t+1}
\left[
K_{j,b}^{\rm sup,(t)}
-
K_{j,b}^{\rm dem,(t)}
\right].
\]

}
}

Compute the price-variation convergence metric
\[
\Delta
=
\left\|
\boldsymbol{\lambda}^{(t+1)}
-
\boldsymbol{\lambda}^{(t)}
\right\|_2
+
\left\|
\boldsymbol{\mu}^{(t+1)}
-
\boldsymbol{\mu}^{(t)}
\right\|_2 .
\]

Set $t\leftarrow t+1$.\;
}

Set $s_{j,b}^{\star}=s_{j,b}^{(t-1)}$ and
$x_{u,j,b}^{\star}=x_{u,j,b}^{(t-1)}$.\;

\Return{$\{s_{j,b}^{\star}\}$, $\{x_{u,j,b}^{\star}\}$}.\;

\end{algorithm}

Constraint \textbf{C1} ensures that each UE is served by exactly one
BS--beam pair. Furthermore, Constraint \textbf{C2} restricts user association to
active beams, i.e., $x_{u,j,b}=1$ only if $s_{j,b}=1$. Next, Constraint
\textbf{C3} limits the number of simultaneously active beams in each
BS to $N_{\mathrm{RF}}$, the number of available RF chains per BS.
In its turn, Constraint \textbf{C4} defines the load $K_{j,b}$ of each 
pair of BS--beams, as used in the load-aware utility. Finally,
Constraint \textbf{C5} enforces the binary nature of the beam
activation and user association variables.

\textit{Theorem 1:} Problem \textbf{P0} is NP-hard.

\textit{Proof:}
Introduce an auxiliary variable $t_m\geq0$ for each satellite such that
$I_m\leq I_m^{\mathrm{th}}+t_m$. For $\lambda_{\max,m}>0$, maximizing the
objective forces $t_m$ to its smallest feasible value, i.e.,
$t_m=[I_m-I_m^{\mathrm{th}}]_+$. Hence, the objective of
\textbf{P0} can be equivalently written as
$U(\mathbf{x})-\sum_{m\in\mathcal{M}}\lambda_{\max,m} t_m$. We reduce the $0$--$1$ knapsack problem to a restricted instance of
\textbf{P0}. Consider $n$ items with values $v_j>0$, integer weights
$a_j\geq0$, and capacity $W$. Construct \textbf{P0} with
$|\mathcal{J}|=|\mathcal{U}|=n$, $|\mathcal{M}|=1$,
$N_{\mathrm{RF}}=1$, and two beams per BS $j$: an item beam
$b_j^{(1)}$ and a dummy beam $b_j^{(0)}$. Set
$c_{j,j,b_j^{(1)}}=e^{v_j}$, $c_{j,j,b_j^{(0)}}=1$, and
$c_{i,j,b}=\delta=e^{-nV-1}$ for $i\neq j$, where
$V=\max_j v_j$. Moreover, set
$g_{j,b_j^{(1)},1}=a_j$, $g_{j,b_j^{(0)},1}=0$,
$I_1^{\mathrm{th}}=W$, and $\lambda_{\max,1}=nV+1$.

Assigning each UE to its corresponding dummy beam is feasible and
yields objective zero. Any cross-BS association contributes at most
$\log\delta=-nV-1$, while the maximum total positive utility is at most
$nV$; hence, any solution containing a cross-BS association has
negative objective and cannot be optimal. Therefore, each UE is
associated with its corresponding BS, and the selected beam has unit
load, i.e., $K_{j,b}=1$. Let $S$ denote the set of BSs activating
their item beams. The objective then becomes
$\sum_{j\in S}v_j-\lambda_{\max,1} t_1$, subject to
$\sum_{j\in S}a_j\leq W+t_1$. Since the item weights $a_j$ and $W$ are integers, any capacity
violation implies $t_1\geq1$ and therefore yields an objective at most
$nV-(nV+1)<0$, whereas the all-dummy solution achieves zero. Thus,
$t_1=0$ at the optimum, and the problem reduces to
$\max_S\sum_{j\in S}v_j$ subject to
$\sum_{j\in S}a_j\leq W$, which is exactly the $0$--$1$ knapsack
problem. Thus, since the construction is polynomial and the $0$--$1$
knapsack problem is NP-hard, \textbf{P0} is NP-hard.
\hfill$\blacksquare$

\vspace{-0.3cm}
\subsection{DT-STABLE Algorithm Development}
\label{subsec:dt_stable_algorithm}

Since \textbf{P0} is computationally intractable due to NP-hardness, we propose a dual-level solution to P0 to obtain a suboptimal yet efficient solution.  The detailed derivation of the two-level solution is provided in Appendix A. Based on this, Algorithm~\ref{alg:dt_stable} proceeds as follows:

\par\noindent
\hangindent=0.9em\hangafter=1
$\bullet$\hspace{0.35em} \textbf{Lines 1--2: Initialization.}
Interference prices $\lambda_m$ and load prices $\mu_{j,b}$ are
initialized to zero, together with the iteration index and convergence
metric.

\par\noindent
\hangindent=0.9em\hangafter=1
$\bullet$\hspace{0.35em} \textbf{Lines 3--6: Provisional UE association.}
Given the current load prices, each UE selects the candidate BS--beam
pair with the highest load-adjusted rate. This provisional association
provides an estimate of the demand and terrestrial utility associated
with each candidate beam before enforcing the RF-chain constraint.

\par\noindent
\hangindent=0.9em\hangafter=1
$\bullet$\hspace{0.35em} \textbf{Lines 7--13: Beam scoring and activation.}
Each BS evaluates its candidate beams using the load-adjusted utilities
of their provisionally associated UEs and the corresponding
satellite-interference price. The occupied beams are ranked according
to their resulting scores and, up to $N_{\mathrm{RF}}$, beams with the
largest scores are activated.

\par\noindent
\hangindent=0.9em\hangafter=1
$\bullet$\hspace{0.35em} \textbf{Lines 14--23: UE reassignment.}
UEs provisionally associated with deactivated beams are reassigned to
the globally active BS--beam pair providing the highest
load-adjusted rate. Since the candidate beams in this step are already
active, the reassignment does not change the aggregate satellite
interference.

\par\noindent
\hangindent=0.9em\hangafter=1
$\bullet$\hspace{0.35em} \textbf{Lines 24--26: Interference-price update.}
For each satellite, the aggregate interference generated by active
beams is compared with the corresponding interference threshold.
A positive interference residual increases the corresponding price,
thereby imposing a larger penalty on highly interfering beams in the
next beam-selection step, whereas a negative residual decreases the
price. The adaptive update strength is controlled by
$\alpha_0$, $c$, $\gamma$, and $\alpha_{\max}$, while the projection
bounds the interference price within its prescribed interval.

\par\noindent
\hangindent=0.9em\hangafter=1
$\bullet$\hspace{0.35em} \textbf{Lines 27--31: Load-price update.}
For each BS--beam pair, the associated UE demand is compared with its
price-induced supply. When the demand exceeds the supply, the load price
increases, reducing attractiveness of the congested beam in the next
association step. Otherwise, the load price decreases to reflect the
lower congestion level.

\par\noindent
\hangindent=0.9em\hangafter=1
$\bullet$\hspace{0.35em} \textbf{Lines 32--35: Convergence and output.}
The convergence metric is evaluated from the changes in interference
and load prices. Iterations terminate when this metric falls below
$\varepsilon$ or when $T_{\max}$ is reached. The resulting beam-activation
and UE-association decisions are then returned.

\textit{Computational Complexity of Algorithm~1:}
In each iteration, provisional UE association and beam-load
evaluation require $\mathcal{O}(UJK)$ operations, while the
interference-aware beam-score computation requires
$\mathcal{O}(JKM)$. Beam selection, interference-price updating,
and load-price updating require $\mathcal{O}(JK\log K)$,
$\mathcal{O}(JKM)$, and $\mathcal{O}(UJK)$ operations,
respectively. Hence, over at most $T_{\max}$ iterations, the overall
complexity is
$\mathcal{O}\!\left(T_{\max}(UJK+JKM+JK\log K)\right)$.

Since $\mathrm{P0}$ is a nonconvex mixed-integer problem, global optimality of Algorithm~\ref{alg:dt_stable} is not guaranteed. Instead, DT-STABLE provides a computationally efficient heuristic that adaptively balances terrestrial capacity and satellite interference protection, particularly when protecting additional satellite links from harmful interference would require a substantial sacrifice in network utility. Section\ref{subsec:complexity_gurobi} evaluates the runtime and optimization performance of DT-STABLE against the Gurobi benchmark.

\vspace{-0.45cm}
\subsection{O-RAN Compliant Multi-Time Scale Optimization} \label{sec:multi_time} 

\begin{algorithm}[!t]
\caption{Multi-Timescale O-RAN DT Resource Optimization}
\label{alg:multi_timescale_dt}
\vspace{-1mm}
\DontPrintSemicolon
\SetAlgoLined
\scriptsize
\SetAlFnt{\scriptsize}
\SetAlCapFnt{\scriptsize}
\SetAlCapNameFnt{\scriptsize}
\SetAlgoNlRelativeSize{-2}
\setlength{\algomargin}{0.1em}
\SetInd{0.05em}{0.15em}
\setlength{\interspacetitleruled}{0pt}
\setlength{\interspacealgoruled}{0pt}
\setlength{\algotitleheightrule}{0.2pt}
\setlength{\algoheightrule}{0.2pt}
\setlength{\textfloatsep}{2pt}
\setlength{\floatsep}{2pt}
\setlength{\intextsep}{2pt}
\setlength{\abovedisplayskip}{1pt}
\setlength{\belowdisplayskip}{1pt}
\setlength{\abovedisplayshortskip}{1pt}
\setlength{\belowdisplayshortskip}{1pt}
\vspace{1mm}

\KwIn{
Beam codebooks $\{\mathcal{B}_j\}_{j\in\mathcal{J}}$
DT coverage map $\mathcal{C}$
DT interference map $\mathcal{G}$
LEO TLE data $\mathcal{T}_{\rm LEO}$
context state $\boldsymbol{\eta}$
interference thresholds $\{I_m^{\rm th}\}$
top-$K_{\rm p}$ probing size $K_{\rm p}$
mismatch threshold $\delta_{\rm mis}$
measured KPM feedback
RF-chain limit $N_{\rm RF}$.
}

\KwOut{
Optimized beam activation $\{s_{j,b}^{\star}\}$ and user association
$\{x_{u,j,b}^{\star}\}$.
}

\textbf{Initialization:}
Set $\mathcal{F}_j\leftarrow\mathcal{B}_j$, $\forall j\in\mathcal{J}$.
$\mathcal{U}_{\rm mis}\leftarrow\emptyset$ and
$\mathcal{K}_u\leftarrow\emptyset$, $\forall u\in\mathcal{U}$.

\textbf{Long-term DT map generation at the Non-RT RIC:}

Collect context information through O1 and update $\boldsymbol{\eta}$.

Update satellite visibility using $\mathcal{T}_{\rm LEO}$ and obtain the
visible satellite set $\mathcal{M}$.

Update the DT coverage map
$\mathcal{C}
=
\{\rho_{n,j,b}\}_{n\in\mathcal{N}_{\rm c},\,j\in\mathcal{J},\,b\in\mathcal{F}_j}$
based on the current context state $\boldsymbol{\eta}$.

Update the DT interference map
$\mathcal{G}
=
\{g_{j,b,m}\}_{j\in\mathcal{J},\,b\in\mathcal{F}_j,\,m\in\mathcal{M}}$
based on satellite visibility and BS beam patterns.

Provide $\mathcal{C}$, $\mathcal{G}$, and $\{\mathcal{F}_j\}$ from the
Non-RT RIC to the Near-RT RIC via the A1 interface.

\textbf{Mid-term DT correction and selective probing at the Near-RT RIC:}

Collect UE locations, serving-cell information, and measured rate KPMs
$\{\hat r_u\}_{u\in\mathcal{U}}$ via the E2 interface.

\ForEach{$u\in\mathcal{U}$}{

Map UE $u$ to its corresponding coverage-map cell $n_u$.

Initialize the UE-specific DT received-power predictions as
$\tilde{\rho}_{u,j,b}\leftarrow\rho_{n_u,j,b}$,
$\forall j\in\mathcal{J}$ and $\forall b\in\mathcal{F}_j$.

Compute the corresponding DT-predicted rates
\[
c_{u,j,b}
\leftarrow
\log_2\!\left(
1+\frac{\tilde{\rho}_{u,j,b}}{\sigma_u^2}
\right),
\quad
\forall j\in\mathcal{J},\;
\forall b\in\mathcal{F}_j .
\]

Obtain the current serving pair
$(j_u^{\rm cur},b_u^{\rm cur})$ and the corresponding
DT-predicted rate
$\tilde r_u\leftarrow
c_{u,j_u^{\rm cur},b_u^{\rm cur}}$.

Compute the relative prediction error
\[
e_u
\leftarrow
\frac{|\hat r_u-\tilde r_u|}
{\max(\tilde r_u,\epsilon_r)},
\]
where $\epsilon_r>0$ is a small numerical constant.

\If{$e_u>\delta_{\rm mis}$}{

$\mathcal{U}_{\rm mis}
\leftarrow
\mathcal{U}_{\rm mis}\cup\{u\}$.

Select the top-$K_{\rm p}$ candidate BS--beam pairs according to the
DT-predicted received power
\[
\mathcal{K}_u
\leftarrow
\operatorname{TopK}_{j\in\mathcal{J},\,b\in\mathcal{F}_j}
\left(\tilde{\rho}_{u,j,b},K_{\rm p}\right).
\]

}
}

Probe the candidate sets
$\{\mathcal{K}_u\}_{u\in\mathcal{U}_{\rm mis}}$
through E2 and obtain the measured received powers
$\{\bar{\rho}_{u,j,b}\}_{(j,b)\in\mathcal{K}_u}$.

Replace the corresponding DT-predicted received-power estimates
for the current scheduling instance,
$\tilde{\rho}_{u,j,b}\leftarrow\bar{\rho}_{u,j,b}$,
$\forall u\in\mathcal{U}_{\rm mis}$ and
$\forall(j,b)\in\mathcal{K}_u$.

Recompute the corresponding achievable rates
\[
c_{u,j,b}
\leftarrow
\log_2\!\left(
1+\frac{\tilde{\rho}_{u,j,b}}{\sigma_u^2}
\right),
\quad
\forall u\in\mathcal{U}_{\rm mis},\;
\forall(j,b)\in\mathcal{K}_u .
\]

\textbf{Short-term scheduling at the Near-RT RIC:}

Run Algorithm~\ref{alg:dt_stable} with inputs
$\{\mathcal{F}_j\}$, $\{c_{u,j,b}\}$, $\{g_{j,b,m}\}$,
$\{I_m^{\rm th}\}$, $\mathcal{U}$, $\mathcal{M}$, and $N_{\rm RF}$.

Obtain $\{s_{j,b}^{\star}\}$ and $\{x_{u,j,b}^{\star}\}$.

Dispatch $\{s_{j,b}^{\star}\}$ and $\{x_{u,j,b}^{\star}\}$ from the
Near-RT RIC to the BSs via the E2 interface.

\Return{$\{s_{j,b}^{\star}\}$, $\{x_{u,j,b}^{\star}\}$}.

\end{algorithm}

The proposed DT-assisted control framework operates across multiple O-RAN timescales to support context-aware NTN--TN coexistence. The non-RT RIC maintains the long-term DT model of the radio environment and uses slowly varying contextual information to construct and update terrestrial CKMs and satellite interference maps. The near-RT RIC, by contrast, processes faster network dynamics, including UE locations, serving-cell information, and measured KPMs collected from the RAN through the E2 interface \cite{Seth2026POST}. This separation enables the DT to combine long-term environmental information with near-RT feedback from the physical network. The DT maintains two CKMs (\textit{C} and \textit{G}) to predict terrestrial beam coverage and interference at satellite receivers. These maps are updated by the non-RT RIC using site-specific context and satellite information, and the resulting feasible beam sets are transferred to the near-RT RIC through the A1 interface. The near-RT xApp then combines the DT predictions with E2 KPM feedback to perform closed-loop resource control.


\begin{table}[t]
\centering
\caption{Simulation, channel-modeling, and DT-STABLE parameters.}
\label{tab:simulation_settings}

\scriptsize
\setlength{\tabcolsep}{4pt}
\renewcommand{\arraystretch}{1.02}

\begin{tabular}{|p{0.53\columnwidth}|p{0.37\columnwidth}|}
\hline
\textbf{Parameter} & \textbf{Value} \\
\hline

\multicolumn{2}{|c|}{\textbf{Network Parameters}} \\
\hline

Terrestrial BSs / UEs
& 19 (~\cite{opencellid}) / 50--300 \\
\hline

LEO satellites / elevation range
& 20 / $25.01^\circ$--$32.46^\circ$ \\
\hline

Deployment area
& $4\times4$~km$^2$ \\
\hline

Carrier frequency / bandwidth
& 28~GHz / 50~MHz~\cite{3gpp_ts38104} \\
\hline

Per-beam transmit power / UE noise figure
& 33~dBm / 7~dB \\
\hline

Monte Carlo runs / snapshots / interval
& 150 / 10 / 1~s \\
\hline

BS / UE height
& 25~m / 1.5~m \\
\hline

UE mobility / speed
& Random waypoint / 0.1--2~m/s \\
\hline

\multicolumn{2}{|c|}{\textbf{Terrestrial BS Antenna Parameters}} \\
\hline

BS array / element spacing
& $4\times4$ URA / $0.5\lambda$ (H/V) \\
\hline

Beam codebook / RF chains
& 16 2-D DFT beams / 4 \\
\hline

Element pattern / downtilt
& 3GPP TR 38.901~\cite{3gpp_tr38901} / $12^\circ$ \\
\hline

Element gain / 3-dB beamwidth
& 8~dBi / $65^\circ$ (H/V) \\
\hline

UE antenna
& Single-element dipole \\
\hline

\multicolumn{2}{|c|}{\textbf{Sionna RT and CKM Parameters}} \\
\hline

Ray-tracing platform
& Sionna RT~\cite{hoydis2023sionnart} \\
\hline

Ray samples / maximum path depth
& $5\times10^{6}$ / 5 \\
\hline

Diffraction / CKM resolution
& Enabled / $5\times5$~m$^2$ \\
\hline

Material models
& ITU-R P.2040 brick, wood, and concrete\\
\hline

\multicolumn{2}{|c|}{\textbf{NTN Channel and Interference Parameters}} \\
\hline

Constellation / ephemeris
& Starlink / CelesTrak TLE~\cite{celestrak} \\
\hline

Minimum visibility elevation
& $25^\circ$ \\
\hline

NTN model / scenario
&  Urban \cite{3gpp_tr38811} \\
\hline

Channel realizations per link
& 50 \\
\hline

Clutter loss (LOS / NLOS)
& 0~dB / elevation- and frequency-dependent \\
\hline

Atmospheric attenuation
& Gaseous, rain, and cloud \\
\hline

Additional losses
& Scintillation / building entry / misc.: 0~dB \\
\hline

Satellite $G/T$ / INR threshold
& 13~dB/K / $-6$~dB$^{\dagger}$ \\
\hline

\multicolumn{2}{|c|}{\textbf{DT-STABLE and Multi-Timescale Parameters}} \\
\hline

Step sizes $(\alpha_0,\beta)$
& $(0.001,\,0.001)$ \\
\hline

Price parameters $(\gamma,c)$
& $(5.0,\,1.2)$ \\
\hline

Caps $(\alpha_{\max},\lambda_{\max,m})$
& $(0.08,\,50)$ \\
\hline

Maximum iterations / tolerance
& 300 / $10^{-3}$ \\
\hline

Minimum UE spectral efficiency
& 0.5~bit/s/Hz \\
\hline

Mismatch threshold / probing size
& $\delta_{\mathrm{mis}}=0.3$ / $K_p=5$ \\
\hline
\multicolumn{2}{|p{0.94\columnwidth}|}{\footnotesize $^{\dagger}$The $-6$~dB 
threshold corresponds to an INR of 25\%, consistent 
with commonly adopted in satellite coexistence 
studies~\cite{kang2024terrestrial}.} \\
\hline
\end{tabular}
\end{table}

\begin{table*}[!t]
    \centering
    \caption{Optimization trade-off and normalized benchmark performance.}
    \label{tab:tradeoff_benchmark}

    \scriptsize
    \renewcommand{\arraystretch}{1.00}
    \setlength{\tabcolsep}{2.6pt}

    \begin{tabular*}{\textwidth}
    {@{\extracolsep{\fill}}|l|c|c|c|c|c|c|}
        \hline

        \multicolumn{7}{|c|}
        {\textbf{(a) Optimization trade-off statistics}}
        \\
        \hline

        \textbf{Approach}
        & \textbf{INR Sat. (\%)}
        & \textbf{Mean INR (dB)}
        & \textbf{Median INR (dB)}
        & \textbf{95th INR (dB)}
        & \textbf{Mean Rate (bps/Hz)}
        & \textbf{95th Rate (bps/Hz)}
        \\
        \hline

        Interference-Optimal
        & 99.99
        & -15.14
        & -14.35
        & -7.94
        & 133.55
        & 206.40
        \\

        DT-STABLE
        & 89.63
        & -13.07
        & -13.09
        & -4.56
        & 210.96
        & 273.47
        \\

        Rate-Optimal
        & 66.99
        & -9.16
        & -9.64
        & 1.84
        & 257.24
        & 304.24
        \\
        \hline
    \end{tabular*}

    \vspace{0.8mm}

    \begin{tabular*}{\textwidth}
    {@{\extracolsep{\fill}}|l|c|c|c|c|c|c|c|}
        \hline

        \multicolumn{8}{|c|}
        {\textbf{(b) Ideal-Point Score Comparison ($U=100$)}}
        \\
        \hline

        \textbf{Metric}
        & \textbf{5G Baseline}
        & \textbf{DT-STABLE}
        & \textbf{NL-Mod.}
        & \textbf{NL-Str.}
        & \textbf{PBS-L}
        & \textbf{PBS-M}
        & \textbf{PBS-H}
        \\
        \hline

        Norm. Rate
        & 1.000
        & 0.740
        & 0.695
        & 0.615
        & 0.703
        & 0.672
        & 0.638
        \\

        Norm. INR Sat.
        & 0.587
        & 0.951
        & 0.970
        & 1.000
        & 0.739
        & 0.867
        & 0.984
        \\

        Norm. Ideal-Point Score
        & 0.708
        & \textbf{0.813}
        & 0.784
        & 0.728
        & 0.720
        & 0.749
        & 0.744
        \\
        \hline
    \end{tabular*}

    \vspace{-1mm}

\end{table*}

A key objective of the proposed multi-timescale design is to avoid exhaustive beam sweeping. Conventional beam sweeping may require testing a large number of BS--beam pairs, which increases measurement overhead and control latency \cite{Khan2023BeamManagement}. By contrast, the DT provides a ranked prediction of promising BS--beam pairs before measurement. When a UE's measured rate is consistent with the DT prediction, the current DT map is reused without triggering additional probing. Beam probing is performed only for UEs whose measured KPMs exhibit a predefined deviation from the DT-predicted KPMs due to residual DT--real-world mismatch. For these UEs, the near-RT xApp selects only the top-$K_{\rm p}$ feasible BS--beam pairs predicted by the DT and refines the corresponding map entries using selective measurements. This DT-guided beam-sweeping procedure reduces the search space while maintaining adaptability to mobility, blockage, and environmental changes. The main steps of Algorithm~\ref{alg:multi_timescale_dt} are summarized as follows:

\par\noindent
\hangindent=0.9em\hangafter=1
$\bullet$\hspace{0.35em} \textbf{Lines 1--4: Initialization and long-term DT update.}
The non-RT RIC collects context information through O1, updates the DT
context state, and determines the visible satellite set from the LEO TLE data.

\par\noindent
\hangindent=0.9em\hangafter=1
$\bullet$\hspace{0.35em} \textbf{Lines 5--6: DT map update.}
The DT updates the coverage map
$\mathcal{C}=\{\rho_{n,j,b}\}$ using the current context state and updates
the interference map $\mathcal{G}=\{g_{j,b,m}\}$ based on the visible
satellites and BS beam patterns.

\par\noindent
\hangindent=0.9em\hangafter=1
$\bullet$\hspace{0.35em} \textbf{Lines 7--9: A1 and E2 information exchange.}
The non-RT RIC provides the updated DT maps and beam sets to the near-RT RIC
through A1. The near-RT RIC collects UE locations, serving-cell information,
and measured rate KPMs through E2.

\par\noindent
\hangindent=0.9em\hangafter=1
$\bullet$\hspace{0.35em} \textbf{Lines 10--20: DT mismatch detection and candidate selection.}
For each UE, the xApp maps its location to the corresponding coverage-map
cell and retrieves the DT-predicted received power for each candidate
BS--beam pair. The achievable rates are then computed. The measured rate
of the current serving pair is compared with its DT-predicted rate, and if
the relative error exceeds $\delta_{\rm mis}$, the UE is added to
$\mathcal{U}_{\rm mis}$. Then, the top-$K_{\rm p}$ candidate BS--beam pairs are selected
according to their DT-predicted received powers.

\par\noindent
\hangindent=0.9em\hangafter=1
$\bullet$\hspace{0.35em} \textbf{Lines 21--23: Selective probing and DT prediction correction.}
For UEs in $\mathcal{U}_{\rm mis}$, only the selected candidate BS--beam
pairs are probed. The measured received powers replace the corresponding
DT-predicted received-power estimates for the current scheduling instance,
and the associated achievable rates are recomputed.

\par\noindent
\hangindent=0.9em\hangafter=1
$\bullet$\hspace{0.35em} \textbf{Lines 24--28: DT-STABLE scheduling and control.}
The resulting achievable rates, interference couplings, feasible beam sets,
and INR thresholds are provided to the DT-STABLE xApp, which executes
Algorithm~\ref{alg:dt_stable}. The resulting beam-activation and UE-association
decisions are sent to the BSs through the E2 interface.
\begin{figure}[!t]
    \centering

    \begin{subfigure}{0.48\columnwidth}
        \centering
        \includegraphics[width=\linewidth]{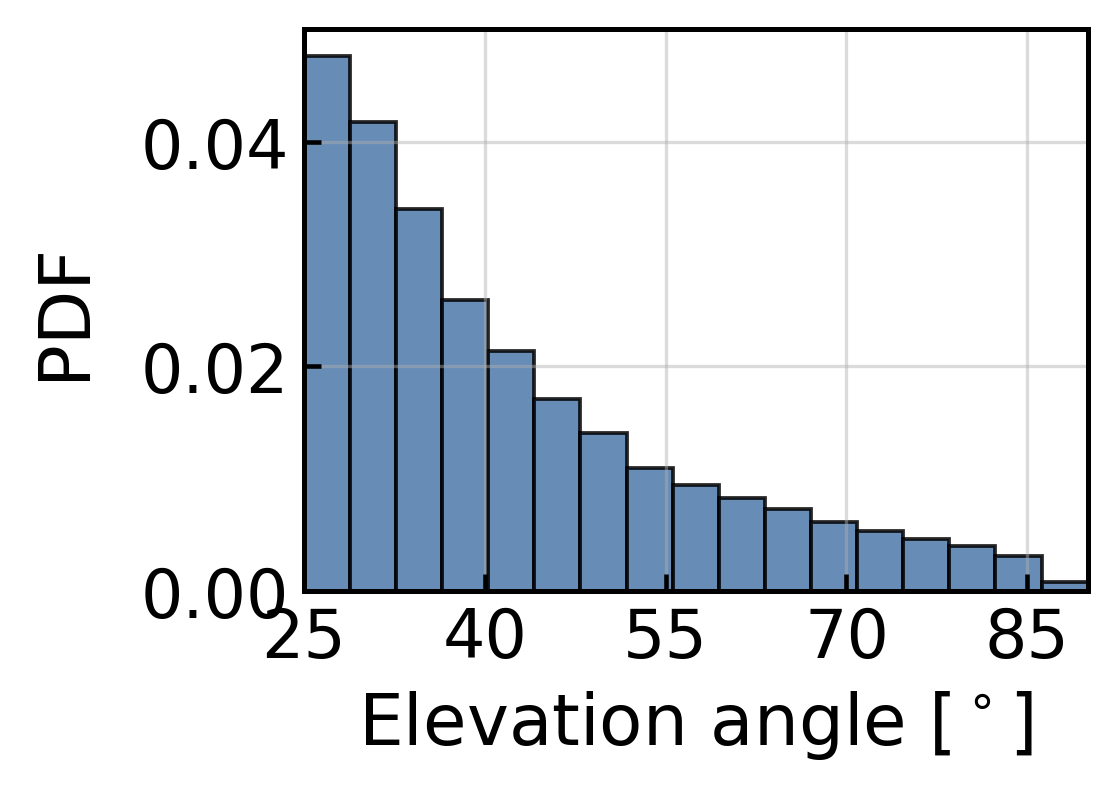}
        \caption{Elevation-angle distribution of visible satellites.}
        \label{fig:satellite_data_analysis}
    \end{subfigure}
    \hfill
    \begin{subfigure}{0.48\columnwidth}
        \centering
        \includegraphics[width=\linewidth]{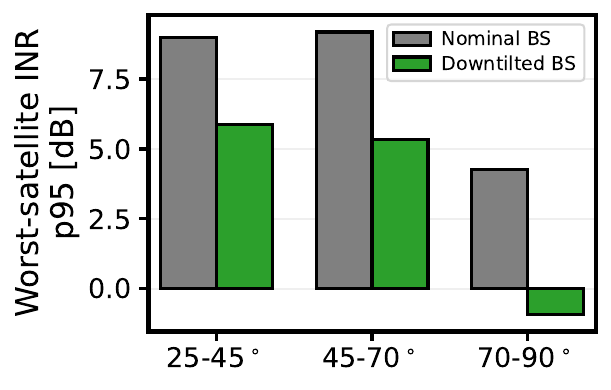}
        \caption{Impact of MBS antenna sidelobes on the worst-satellite INR.}
        \label{fig:downtilt_effect_worst_satellite}
    \end{subfigure}

    \caption{Satellite visibility and sidelobe-induced interference characteristics in the considered coexistence scenario.}
    \label{fig:satellite_and_sidelobe_analysis}
    \vspace{-0.1cm}
\end{figure}

\begin{figure}[!t]
    \centering

    \begin{minipage}[t]{0.49\columnwidth}
        \vspace{0pt}
        \centering
        \includegraphics[
            width=\linewidth,
            trim=0 0 0 0,
            clip
        ]{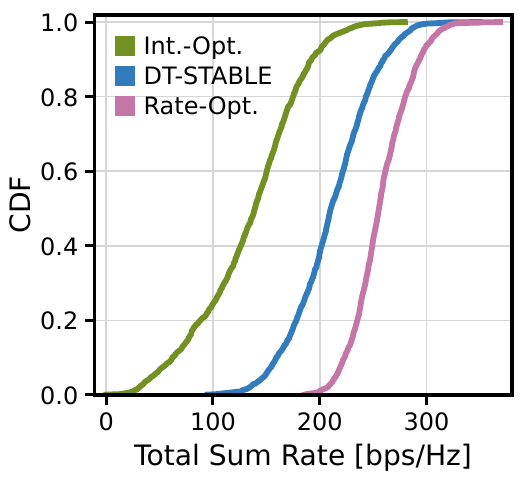}

        \vspace{-2mm}
        {\scriptsize (a)}
    \end{minipage}%
    \hfill%
    \begin{minipage}[t]{0.49\columnwidth}
        \vspace{0pt}
        \centering
        \includegraphics[
            width=\linewidth,
            trim=0 0 0 0,
            clip
        ]{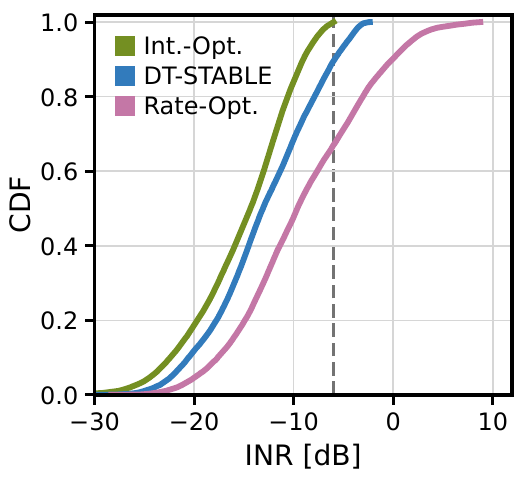}

        \vspace{-2mm}
        {\scriptsize (b)}
    \end{minipage}

    \vspace{-1mm}

    \caption{CDF comparison of the benchmark schemes:
    (a) terrestrial sum rate and (b) satellite interference.}
    \label{fig:cdf_benchmark}

    \vspace{-2mm}
\end{figure}

\vspace{-0.4cm}
\section{Performance Evaluation}
\label{sec:performance_eval}
\vspace{-0.25cm}
\subsection{Simulation Setting}

The proposed DT-STABLE framework was evaluated using a synchronized CDT--physical-network simulation environment. The CDT provided site-specific
radio information to support decision making, while the physical-network layer evaluated the resulting beam activation and UE association decisions. Robustness of DT-STABLE against practical impairments is evaluated in Section~\ref{subsec:robustness_evaluation}. The simulation, channel-modeling, Sionna RT/CKM, and DT-STABLE multi-timescale control parameters are summarized in Table~\ref{tab:simulation_settings}. We considered NTN--TN spectrum coexistence, where terrestrial downlink transmissions share resources with a LEO satellite system. The site-specific environment was modeled as a DT of a $4\times4$~km$^2$ region near Québec City, Canada. The TN comprised a total of 19 macro O-RUs/MBSs whose geographic coordinates were obtained from the OpenCellID database~\cite{opencellid}. Following the 3GPP suburban macro assumptions, each macro O-RU had a height of 25~m and a 1000~m service region. Each O-RU was equipped with a half-wavelength-spaced directional $4\times4$ URA and a finite DFT codebook of 16 candidate beams.

%
%

\begin{figure*}[!t]
    \centering


    \begin{minipage}[t]{0.3333\textwidth}
        \vspace{0pt}
        \centering

        \includegraphics[
            width=\linewidth
        ]{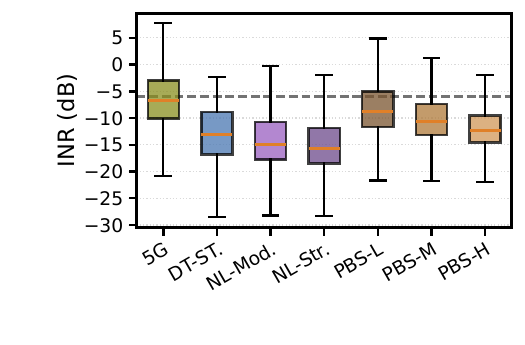}

        \vspace{-8mm}
        {\small (a)}
    \end{minipage}%
    \begin{minipage}[t]{0.3333\textwidth}
        \vspace{0pt}
        \centering

        \includegraphics[
            width=\linewidth
        ]{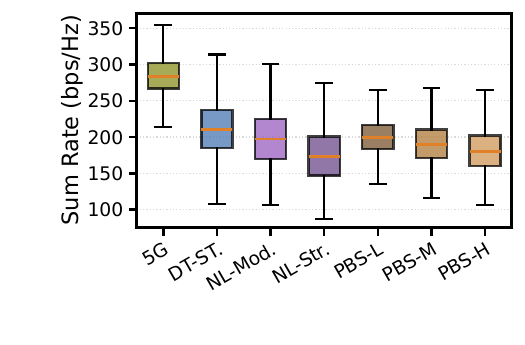}

        \vspace{-8mm}
        {\small (b)}
    \end{minipage}%
    \begin{minipage}[t]{0.3333\textwidth}
        \vspace{0pt}
        \centering

        \includegraphics[
            width=\linewidth
        ]{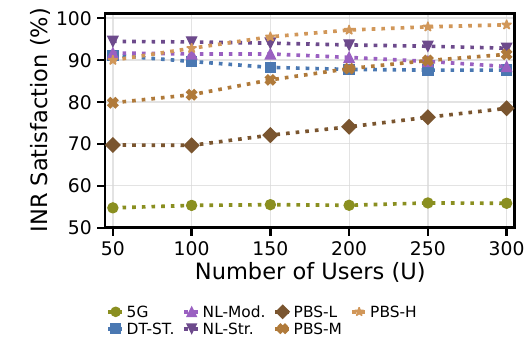}
    \end{minipage}


    \begin{minipage}[t]{0.6666\textwidth}
        \vspace{-1.5mm}
        \centering
        \captionof{figure}{
            Benchmark distributions:
            (a) satellite INR and
            (b) terrestrial sum rate.
        }
        \label{fig:benchmark_distributions}
    \end{minipage}%
    \begin{minipage}[t]{0.3333\textwidth}
        \vspace{-1mm}
        \centering
        \captionof{figure}{
            INR satisfaction vs. the number of users.
        }
        \label{fig:inr_satisfaction}
    \end{minipage}




    \begin{minipage}[t]{0.3333\textwidth}
        \vspace{0pt}
        \centering

        \includegraphics[
            width=\linewidth
        ]{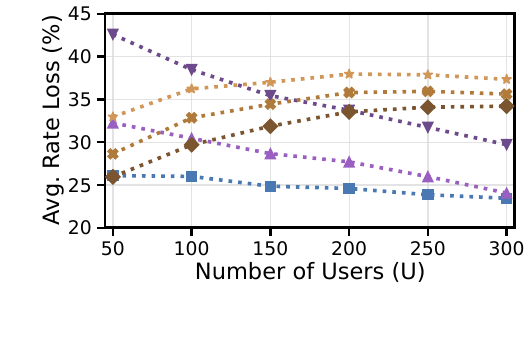}
    \end{minipage}%
    \begin{minipage}[t]{0.3333\textwidth}
        \vspace{0pt}
        \centering

        \includegraphics[
            width=\linewidth
        ]{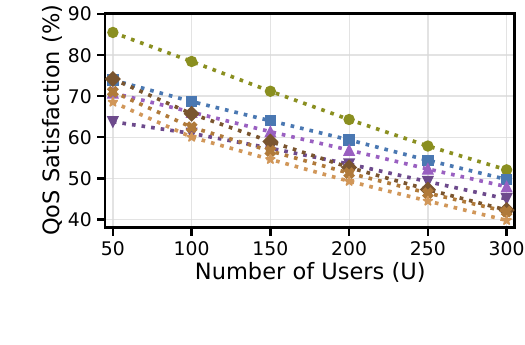}
    \end{minipage}%
    \begin{minipage}[t]{0.3333\textwidth}
        \vspace{0pt}
        \centering

        \includegraphics[
            width=\linewidth
        ]{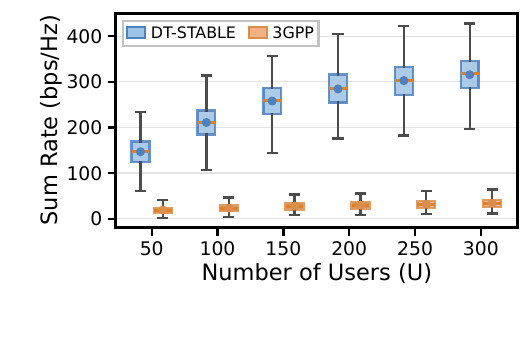}
    \end{minipage}


    \begin{minipage}[t]{0.3333\textwidth}
        \vspace{-5mm}
        \centering
        \captionof{figure}{
            Average user-rate loss vs. the number of users.
        }
        \label{fig:rate_loss}
    \end{minipage}%
    \begin{minipage}[t]{0.3333\textwidth}
        \vspace{-5mm}
        \centering
        \captionof{figure}{
            QoS satisfaction vs. the number of users.
        }
        \label{fig:qos_satisfaction}
    \end{minipage}%
    \begin{minipage}[t]{0.3333\textwidth}
        \vspace{-5mm}
        \centering
        \captionof{figure}{
            DT-STABLE vs. 3GPP sum rate.
        }
        \label{fig:dt_3gpp_sumrate}
    \end{minipage}
\vspace{-0.8cm}
\end{figure*}

For the considered LEO constellation, satellite positions were derived from TLE data obtained from CelesTrak~\cite{celestrak}. Satellite visibility across the deployment area was determined using a minimum elevation threshold of $25^\circ$. Under this condition, Fig.~\ref{fig:satellite_data_analysis} shows that the elevation-angle distribution of the visible Starlink satellites is concentrated toward the lower-elevation range. Fig.~\ref{fig:downtilt_effect_worst_satellite} further examines the impact of elevation angle on terrestrial interference by incorporating the MBS antenna response into the calculation of the worst-case INR. While antenna downtilt reduced the p95 INR by up to $5.19$~dB at elevations above $70^\circ$, the reduction decreased to $3.09$~dB in the $25^\circ$--$45^\circ$ range, leaving the residual p95 INR above the $-6$~dB protection threshold. Thus, low-elevation satellites constitute the most critical coexistence regime in the considered scenario, as they are both more prevalent in the visibility set and less effectively protected by antenna downtilt. To provide a conservative worst-case assessment of DT-STABLE, the ensuing analysis considered visible satellites with the lowest elevation angles at each observation.
\vspace{-0.4cm}
\subsection{Optimization Strategies}
\label{subsec:optimization_strategies}
By tuning the adaptive interference-price parameters $(\gamma,c,\alpha_{\max})$ 
in Algorithm~\ref{alg:dt_stable}, which control how strongly the interference-prices
 respond to constraint violations, the proposed algorithm realizes three 
operating points: \emph{interference} optimal, \emph{balanced}, and 
rate \emph{optimal}. The interference-optimal point prioritizes satellite 
protection, whereas the rate-optimal point prioritizes terrestrial throughput.

Fig.~\ref{fig:cdf_benchmark}(b), Fig.~\ref{fig:cdf_benchmark}(a), and Table~\ref{tab:tradeoff_benchmark}(a) characterize 
the rate--interference trade-off among the three operating points. 
Fig.~\ref{fig:cdf_benchmark}(b) presents the INR CDF, where the vertical dashed line 
marks the target threshold of $-6$~dB, while 
Table~\ref{tab:tradeoff_benchmark}(a) summarizes the corresponding INR 
satisfaction and terrestrial sum-rate performance. For operating scheme $a$, the INR satisfaction ratio was defined as
$\eta_{a}^{\mathrm{INR}}=\frac{100\%}{RSM}\sum_{r=1}^{R}\sum_{s=1}^{S}\sum_{m=1}^{M}
\mathbf{1}[\mathrm{INR}_{a,m}^{(r,s)}\leq \mathrm{INR}_{\mathrm{th}}]$,
where $R$, $S$, and $M$ denoted the numbers of simulation runs, snapshots,
and evaluated satellites, respectively. The interference-optimal configuration achieved the highest INR satisfaction, $99.99\%$, but the lowest mean system sum-rate, $133.55$~bps/Hz. In other words, nearly every evaluated satellite remained below the INR threshold under this configuration. By contrast, the rate-optimal configuration achieved the highest mean system sum-rate, $257.24$~bps/Hz, while its INR satisfaction decreased to $66.99\%$. DT-STABLE provided a balanced operating point, achieving $89.63\%$ INR satisfaction and $210.96$~bps/Hz. Compared to the rate-optimal setting, it improved INR satisfaction by $22.64$ percentage points, while increasing the sum-rate by approximately $1.58\times$ relative to the interference-optimal setting. The balanced point, referred to as DT-STABLE, was used throughout the remainder of the evaluation, as it provides an effective balance between terrestrial throughput and satellite interference protection.
\setlength{\dbltextfloatsep}{5pt}

\vspace{-0.4cm}
\subsection{Benchmark Schemes}

We compared DT-STABLE with the three benchmark schemes in terms of satellite INR, terrestrial sum rate, average user-rate loss, and QoS satisfaction. To capture different levels of INR-constraint enforcement, the no-load-aware scheme was evaluated under moderate and strict filtering levels, while the per-BS beam selection scheme was assessed under low, medium, and high local beam-penalty levels.
\par\noindent
\hangindent=0.9em\hangafter=1
$\bullet$\hspace{0.35em} \textbf{Baseline 1: 5G Baseline.} This scheme follows the conventional 3GPP beam management procedure, where each BS exhaustively sweeps all candidate DFT beams and each user is assigned to the BS–beam pair with the highest channel gain. Satellite interference was not considered, making this a standard-compliant, but coexistence-unaware benchmark that served as the terrestrial-rate upper bound.

\par\noindent
\hangindent=0.9em\hangafter=1
$\bullet$\hspace{0.35em} \textbf{Baseline 2: No-Load-Aware  (NL) Scheme.} This benchmark does not account for load-dependent bandwidth sharing among users served by the same BS--beam pair. Thus, multiple users can be assigned to the same beam without an explicit load penalty. Two variants, moderate and strict, were considered by changing the weight of the satellite INR term in the optimizer.
\begin{figure*}[!t]
\centering

\begin{minipage}[t]{0.50\textwidth}
\vspace{0pt}
\centering

\includegraphics[
    width=\linewidth
]{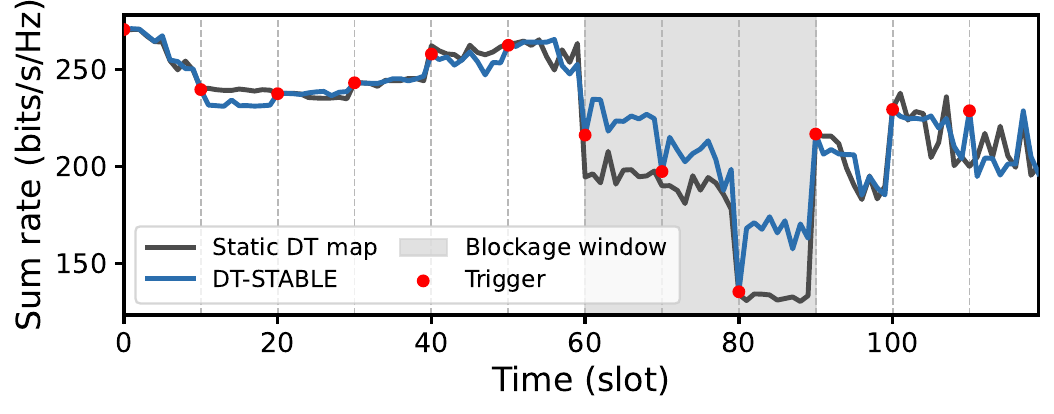}
\vspace{-0.78cm}
\captionof{figure}{Sum-rate comparison during temporary blockage.}
\label{fig:tf_beam}

\end{minipage}
\hfill
%
\begin{minipage}[t]{0.49\textwidth}
\vspace{0pt}
\centering

\begin{minipage}[t]{0.53\linewidth}
\vspace{0pt}
\centering

\includegraphics[
    width=\linewidth
]{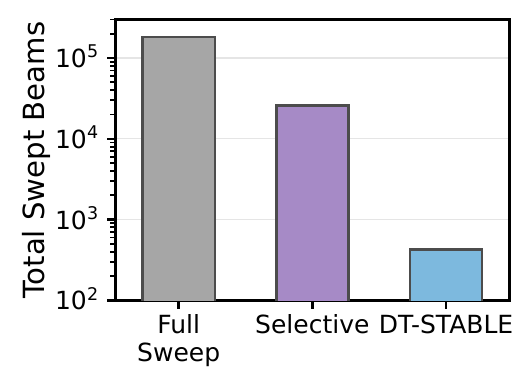}
\vspace{-0.8cm}
\captionof{figure}{Total number of swept beams.}
\label{fig:total_swept_beams}

\end{minipage}
\hfill
%
\begin{minipage}[t]{0.44\linewidth}
\vspace{0pt}
\centering

\captionsetup{font=scriptsize}
\captionof{table}{Recovered rate versus beam sweeping budget.}
\label{tab:k_sweep_best_rate}

\vspace{1mm}

\scriptsize
\setlength{\tabcolsep}{2.5pt}
\renewcommand{\arraystretch}{0.90}

\begin{tabular}{|c|c|}
\hline
Swept beams, $K_{\rm p}$ & Normalized rate \\
\hline
1  & 0.670 \\
\hline
2  & 0.793 \\
\hline
3  & 0.841 \\
\hline
5  & 0.867 \\
\hline
8  & 0.919 \\
\hline
16 & 1.000 \\
\hline
\end{tabular}

\end{minipage}

\end{minipage}

\vspace{-2mm}
\end{figure*}

\par\noindent
\hangindent=0.9em\hangafter=1
$\bullet$\hspace{0.35em} \textbf{Baseline 3: Per-BS Beam (PBS) Selection.} 
This benchmark performs interference-aware beam selection locally at each BS, inspired by \cite{kang2024terrestrial}. 
PBS-Low, PBS-Medium, and PBS-High denote low, medium, and high local beam-penalty levels, respectively, with higher levels providing more conservative satellite protection. 
Unlike DT-STABLE, PBS does not coordinate aggregate interference across BSs or balance the load across active beams.

Fig.~\ref{fig:benchmark_distributions}(a) presents the aggregate INR distribution for $U=100$. The 5G baseline yielded a mean INR of $-6.53$~dB with only 55.32\% satisfaction, showing that rate-driven allocation alone cannot ensure NTN--TN coexistence. DT-STABLE reduced the mean INR to $-13.07$~dB with 89.63\% satisfaction, achieving INR control comparable to the NL schemes while avoiding excessive beam suppression. By contrast, PBS-Low under-controls INR, whereas PBS-High provides more conservative interference protection.

Fig.~\ref{fig:benchmark_distributions}(b) shows the corresponding sum-rate distribution. Although the 5G baseline achieved the highest rate because it is unconstrained by INR, DT-STABLE reached 210.96~bps/Hz, outperforming NL-Moderate/Strict, PBS-Medium/High, and the other schemes providing comparable INR protection. This rate--interference trade-off is also reflected in Table~\ref{tab:tradeoff_benchmark}(b).

For each scheme $a$, the rate and INR-satisfaction performance were normalized by their respective maxima as
$R_a^{\mathrm{norm}}=\bar{R}_a/\max_{q\in\mathcal{A}}\bar{R}_q$
and
$P_a^{\mathrm{norm}}=\eta_a^{\mathrm{INR}}/\max_{q\in\mathcal{A}}\eta_q^{\mathrm{INR}}$,
so that both ratios are dimensionless and $(1,1)$ represents the theoretically perfect solution. We then defined the normalized ideal-point score
$\mathcal{S}_a = 1-\sqrt{(1-R_a^{\mathrm{norm}})^2+(1-P_a^{\mathrm{norm}})^2}/\sqrt{2}$,
which measures each scheme's Euclidean distance to this ideal point, with
$\mathcal{S}_a\in[0,1]$ and larger values indicating closer proximity. DT-STABLE attained the highest score, $\mathcal{S}_a=0.813$, indicating the closest proximity to the ideal point among the evaluated benchmarks.

Fig.~\ref{fig:inr_satisfaction} shows a similar trend across user densities. The 5G baseline achieved approximately 55\% INR satisfaction for all $U$, while the PBS schemes improved satisfaction through higher penalty levels at the expense of capacity. The NL schemes also became slightly less effective as the number of users increased; for example, the moderate setting decreased from 91.65\% to 88.43\%. DT-STABLE maintained a more balanced behavior, with INR satisfaction decreasing from 91.09\% at $U=50$ to 87.56\% at $U=300$, as it jointly incorporated DT-predicted rates, INR prices, and load prices.

Fig.~\ref{fig:rate_loss} further illustrates this trade-off. The more conservative NL schemes provided stronger interference protection but incurred larger user-rate loss relative to the 5G baseline at lower user densities, whereas the PBS schemes exhibited increasing rate loss as the number of users increased. At $U=100$, DT-STABLE achieved 89.63\% INR satisfaction with a 25.99\% rate loss relative to the 5G baseline and maintained higher INR satisfaction than PBS-Low across all evaluated user densities.

\vspace{-0.1cm}
Fig.~\ref{fig:qos_satisfaction} reports QoS satisfaction across user densities, defined as the percentage of UEs whose achieved spectral efficiency meets the minimum requirement $r_{\min}=0.5$~bit/s/Hz. DT-STABLE's QoS satisfaction decreased from 73.91\% to 49.73\% as $U$ increased from 50 to 300, while remaining more stable than the other INR-aware schemes. At $U=100$, it achieved 68.67\% QoS satisfaction, outperforming the PBS schemes and remaining competitive with the NL schemes. These results indicate that load pricing helps limit excessive user concentration on active BS--beam pairs.
\vspace{-0.4cm}
\subsection{Advantages of Digital Twin Predicted Channel}

We evaluated the effect of the DT-predicted, site-specific channel model by comparing it to a conventional 3GPP urban macrocell (UMa) baseline. The DT channel maps were generated using Sionna RT in a reconstructed 3D environment, incorporating actual BS locations, terrain and building geometry, material-dependent propagation, and multi-antenna beam patterns. This approach enabled the DT model to capture channel variations due to blockage, reflection, diffraction, and scattering.
Fig.~\ref{fig:dt_3gpp_sumrate} compares the total sum rate achieved with the DT-predicted channel and the statistical 3GPP baseline at different user densities. The DT-based model consistently outperformed the 3GPP baseline. With an increase of the number of users from 50 to 300, the average sum rate with the DT-predicted channel rose from 146.85 to 315.64~bps/Hz, whereas the 3GPP baseline increased only from 25.49 to 43.45~bps/Hz. The median sum rate followed a similar trend, with the DT-based model achieving 147.07--318.07~bps/Hz as compared to 24.47--42.87~bps/Hz for the 3GPP abstraction. Moreover, the relative gain of the DT-based model widened from approximately $5.8\times$ to $7.3\times$ as user density increased.
These results demonstrated that channel fidelity is essential for site-specific resource allocation. Unlike statistical models, the DT-based channel model captured physics-grounded, deployment-specific RF conditions and provided link-quality estimates that closely reflected the actual BS--UE environment, thus enabling more effective beam selection and user association.
%
%

\begin{figure*}[!t]
    \centering


    \begin{minipage}[t]{0.3333\textwidth}
        \vspace{0pt}
        \centering

        \includegraphics[
            width=0.88\linewidth
        ]{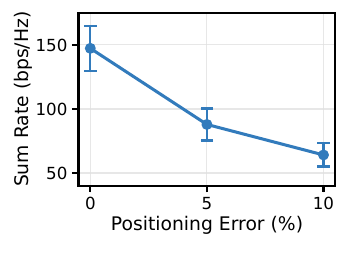}

        \vspace{-5.0mm}
        {\small (a)}
    \end{minipage}%
    \begin{minipage}[t]{0.3333\textwidth}
        \vspace{0pt}
        \centering

        \includegraphics[
            width=0.88\linewidth
        ]{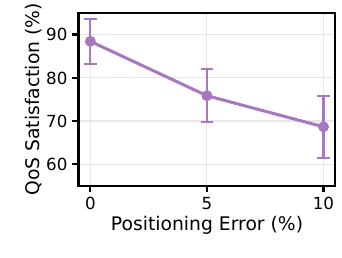}

        \vspace{-5.0mm}
        {\small (b)}
    \end{minipage}%
    \begin{minipage}[t]{0.3333\textwidth}
        \vspace{0pt}
        \centering

        \includegraphics[
            width=0.88\linewidth
        ]{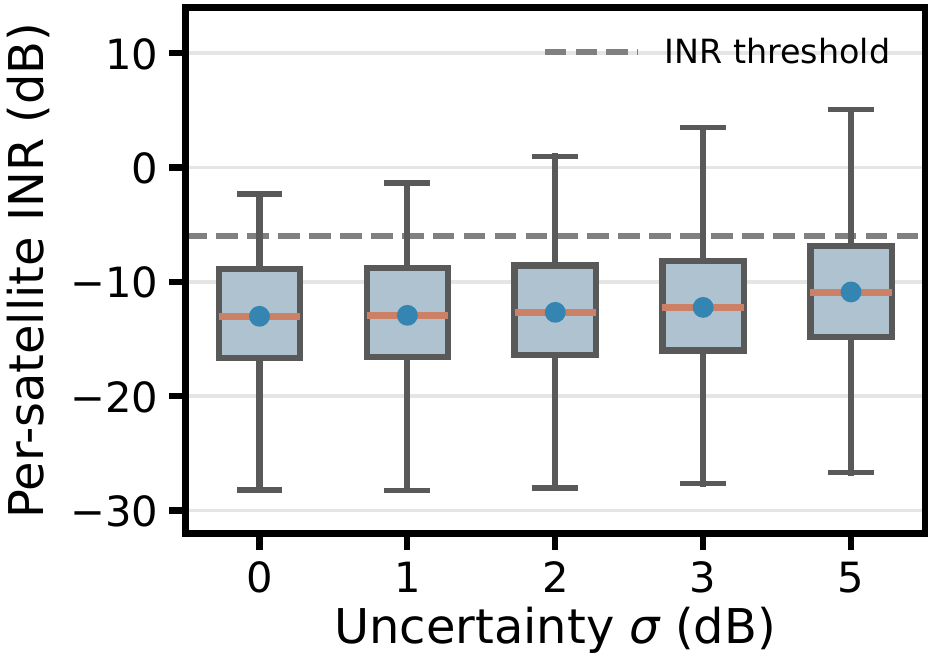}
    \end{minipage}


    \begin{minipage}[t]{0.6666\textwidth}
        \centering

        \captionof{figure}{
        Impact of UE positioning error:
        (a) sum rate and (b) QoS satisfaction.
        }
        \label{fig:positioning_error}
    \end{minipage}%
    \begin{minipage}[t]{0.3333\textwidth}
        \centering
        \vspace{-0.2cm}
        \captionof{figure}{
        Impact of interference-map uncertainty on INR.
        }
        \label{fig:interference_uncertainty}
    \end{minipage}


    \vspace{-0.2cm}

    \begin{minipage}[t]{0.3333\textwidth}
        \vspace{0pt}
        \centering

        \includegraphics[
            width=0.88\linewidth
        ]{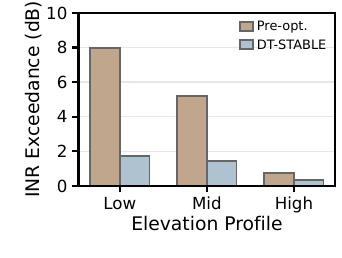}
    \end{minipage}%
    \begin{minipage}[t]{0.3333\textwidth}
        \vspace{0pt}
        \centering

        \includegraphics[
            width=0.88\linewidth
        ]{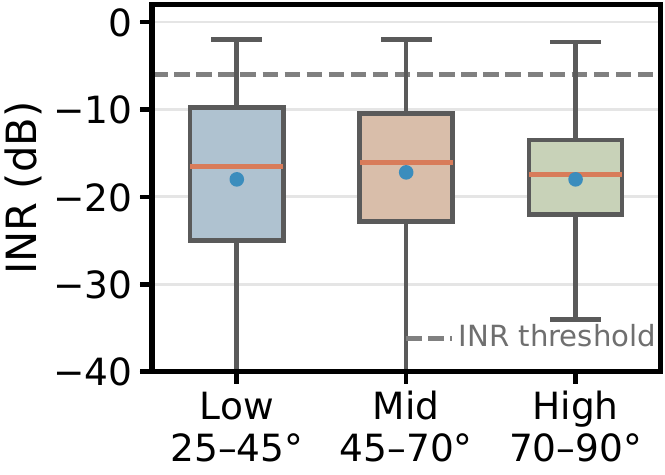}
    \end{minipage}%
    \begin{minipage}[t]{0.3333\textwidth}
        \vspace{0pt}
        \centering

        \includegraphics[
            width=0.88\linewidth
        ]{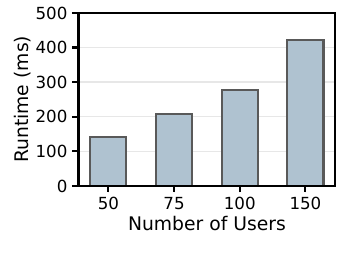}
    \end{minipage}


    \begin{minipage}[t]{0.3333\textwidth}
        \vspace{-1.0mm}
        \centering

        \captionof{figure}{
        Mean INR exceedance vs. elevation profiles.
        }
        \label{fig:elevation_exceedance}
    \end{minipage}%
    \begin{minipage}[t]{0.3333\textwidth}
        \vspace{-1.0mm}
        \centering

        \captionof{figure}{
        INR distribution across elevation profiles.
        }
        \label{fig:elevation_inr}
    \end{minipage}%
    \begin{minipage}[t]{0.3333\textwidth}
        \vspace{-1.0mm}
        \centering

        \captionof{figure}{
        DT-STABLE runtime versus user density.
        }
        \label{fig:runtime}
    \end{minipage}
\vspace{-0.3cm}
\end{figure*}
\vspace{-0.3cm}
\subsection{Advantages of Multi-time Scale Digital Twin Assisted Resource Management}

To assess DT-assisted beam management under short-term radio variations, we consider a 120-slot time-domain experiment based on the 28-GHz DT environment. The channel state is held fixed over each 10-slot feedback interval and updated at the beginning of the next interval, consistent with the adopted DT-refinement timescale. During slots 60--90, a temporary mixed-blockage event is introduced for 30\% of the UEs. Among the affected UEs, a subset experiences complete blockage of the scheduled links, whereas the remaining links undergo partial attenuation. This creates a controlled mismatch between the offline CKM and instantaneous radio conditions. The static DT continues to rely on the offline CKM, whereas DT-STABLE uses measurement feedback to identify and correct inaccurate link predictions. At each feedback instant, the DT-predicted rate of the serving link is compared with the UE measurement. When the relative prediction error exceeds the predefined mismatch threshold, DT-STABLE triggers selective probing of DT-ranked candidate beams and updates the corresponding CKM information before resource allocation.

Fig.~\ref{fig:tf_beam} shows the temporal sum-rate evolution of the static DT and DT-STABLE under the blockage experiment. The setup considers 50 UEs, 19 BSs, 16 candidate beams per BS, and feedback evaluation every 10 slots. Prior to the blockage interval, the two schemes exhibit comparable performance because the offline CKM remains consistent with the channel conditions. Once the blockage is introduced, the static DT cannot account for the resulting channel mismatch and continues to make resource-allocation decisions from outdated link information. In contrast, DT-STABLE incorporates the feedback-driven CKM refinement in Algorithm~\ref{alg:multi_timescale_dt} and adapts the affected beam and association decisions accordingly. Over the blockage interval, the average sum rate increases from 172.40~bits/s/Hz with the static DT to 197.91~bits/s/Hz with DT-STABLE, corresponding to a 14.8\% improvement.

Fig.~\ref{fig:total_swept_beams} quantifies the corresponding beam-measurement overhead. A full sweeping strategy evaluates all BS--beam candidates for every UE at each feedback instant, requiring 182,400 measurements. Selective exhaustive sweeping limits these measurements to links with a detected prediction mismatch, reducing the total to 32,528 measurements. DT-STABLE further restricts the search to the top-$K_{\rm p}$ candidates identified from the DT. With $K_{\rm p}=5$, the 107 detected mismatch events require only 535 beam measurements, corresponding to reductions of 99.71\% and 98.36\% relative to full and selective exhaustive sweeping, respectively. Table~\ref{tab:k_sweep_best_rate} further evaluates the quality of the DT-guided beam ranking under different probing budgets. For each mismatch event, the highest rate among the top-$K_{\rm p}$ CKM-ranked beams of the serving BS is normalized by the highest rate from exhaustive probing of all 16 beams, and the resulting ratios are averaged over all mismatch events. With $K_{\rm p}=8$, the DT-guided candidate set achieves 91.9\% of the full-probing rate on average, demonstrating that measurement overhead can be substantially reduced while retaining most of the rate available from exhaustive beam probing.
\vspace{-0.4cm}
\subsection{Robustness Evaluation}
\label{subsec:robustness_evaluation}
We further evaluated the robustness of DT-STABLE against practical impairments. In this evaluation, the scheduler operated with mismatched DT information, while the resulting beam activation and user association decisions were validated over the real-world layer. We examined the following two main sources of DT mismatch: errors in UE positioning and uncertainty in the interference map. We also analyzed how the satellite elevation angle influences interference and coexistence performance by altering the BS–satellite geometry.

\textit{1) Impact of Imperfect UE Positioning Estimation:}
Fig.~\ref{fig:positioning_error} shows the impact of UE positioning errors on DT-STABLE performance for 50 users. Since the CKM was queried based on UE locations, an error in the estimated position may cause the scheduler to use channel information that is not fully aligned with the actual UE location. This can lead to less effective BS--beam associations and reduce the achieved terrestrial performance. With accurate UE positioning, DT-STABLE achieved a mean total sum rate of 147.33~bps/Hz and a QoS satisfaction ratio of 88.4\%. When the positioning error increased to 5\%, these values decreased to 87.88~bps/Hz and 75.8\%, respectively. At 10\% positioning error, the mean total sum rate and QoS satisfaction ratio further decreased to 64.06~bps/Hz and 68.6\%, respectively. Overall, the total sum rate decreased by 56.5\%, whereas the QoS satisfaction ratio decreased by 19.8 percentage points. This indicates that positioning errors primarily degraded the achievable rate, while their effect on the ability to satisfy the minimum user-rate requirement was comparatively smaller. These results highlight the importance of accurate spatial information for the performance of the DT-assisted scheduler.

\textit{2) Impact of Interference Prediction Uncertainty:} Fig.~\ref{fig:interference_uncertainty} evaluates DT-STABLE under the relaxed Assumption~A2 for 100 users, where $\sigma_I$ denotes the standard deviation of the prediction error applied to the DT beam-to-satellite coupling coefficients. The scheduler operates using the nominal interference map, while the actual couplings are modeled as $g_{j,k,m}^{\mathrm{true}}=g_{j,k,m}^{\mathrm{DT}}10^{\epsilon_{j,k,m}/10}$, with $\epsilon_{j,k,m}\sim\mathcal{N}(0,\sigma_I^2)$. Thus, $\sigma_I=0$~dB represents an exact interference map, while larger values introduce increasing coupling uncertainty. The terrestrial channel rates are kept unchanged to isolate this effect. As $\sigma_I$ increases, the INR distribution shifts toward higher values. At $\sigma_I=0$~dB, the mean and median INR are $-13.01$~dB and $-13.05$~dB, respectively. At $\sigma_I=5$~dB, they increase to $-10.88$~dB and $-10.92$~dB, while the INR satisfaction ratio decreases from 89.66\% to 79.15\%, demonstrating the impact of interference-map uncertainty on satellite protection.

\textit{3) Impact of Elevation Angle:} We next evaluated how the satellite elevation angle affects DT-STABLE's worst-case INR performance. This is relevant because BS--satellite interference depends on the propagation geometry and the MBS antenna response toward the satellite direction. Fig.~\ref{fig:elevation_exceedance} shows that, before optimization, the worst-satellite INR excedence is elevation dependent. The  worst-case INR for all three elevation profiles exceeded the target INR threshold, with mean exceedances of 7.2~dB, 5.1~dB, and 0.8~dB for the low- ($25^\circ$--$45^\circ$), mid- ($45^\circ$--$70^\circ$), and high-elevation ($70^\circ$--$90^\circ$) cases, respectively. After DT-STABLE scheduling, these exceedances were significantly reduced to 1.8~dB, 1.5~dB, and 0.2~dB, corresponding to reductions of 75.0\%, 70.6\%, and 75.0\%, respectively.

Fig.~\ref{fig:elevation_inr} further shows the post-scheduling per-satellite INR distributions for the three elevation profiles. The median INR remained comparable across the low-, mid-, and high-elevation cases, with values of $-16.47$~dB, $-16.17$~dB, and $-17.55$~dB, respectively. In contrast, the upper tail of the INR distribution improved with increasing elevation, as the 95th-percentile INR decreased from $-4.54$~dB at low elevation to $-5.11$~dB at mid elevation and $-8.52$~dB at high elevation. Accordingly, the INR satisfaction ratio increased from 89.63\% at low elevation to 92.47\% and 98.50\% at mid and high elevation, respectively, indicating that DT-STABLE maintained effective satellite protection across all elevation profiles, with lower upper-tail interference at higher elevation profiles.
\vspace{-0.32cm}

\subsection{Computational Complexity and Gurobi Benchmarking}
\label{subsec:complexity_gurobi}

Since DT-STABLE was executed by the near-RT resource-allocation xApp, its runtime is critical for practical deployment. Fig.~\ref{fig:runtime} reports computation time for different UE populations. As expected, the runtime increased with the number of UEs because a larger user set expands the BS--beam association space. However, it remained below $0.45$~s in all cases, including the 150-UE scenario, indicating compatibility with the near-RT RIC control timescale of 10~ms--1~s.

To evaluate its computational efficiency and solution quality, DT-STABLE was
compared with Gurobi on reduced instances of Problem~$\mathrm{P0}$.
Table~\ref{tab:results} reports the runtime, sum rate per BS, and INR
satisfaction for different BS and UE configurations. Gurobi solves the
mixed-integer problem using branch-and-bound, which searches the
combinatorial beam-activation and UE-association space while enforcing the
constraints of Problem~$\mathrm{P0}$. Therefore, it achieved 100\% INR
satisfaction, but required significantly higher runtime. Gurobi required
$3.26$--$21.21$~s, whereas DT-STABLE needed only
$0.207$--$0.757$~s. DT-STABLE avoids this combinatorial search through
iterative price-based beam selection and user association, achieving much
lower runtime with comparable sum-rate performance, although its INR
satisfaction was up to 30 percentage points lower than that of Gurobi.
For example, DT-STABLE achieved $14.09$ versus $13.38$~bps/Hz/BS for
12 BSs and 200 UEs, and $12.60$ versus $13.00$~bps/Hz/BS for 12 BSs
and 220 UEs. In six of the ten configurations, DT-STABLE attains a slightly
higher sum rate than Gurobi since it does not strictly enforce the INR
constraints for all satellites. These results show that DT-STABLE provides
a practically implementable solution within the near-RT time constraint,
with limited sum-rate and INR-satisfaction degradation compared with exact
optimization.

\begin{table}[!t]
\centering

\captionsetup{font=footnotesize}
\caption{Performance comparison between Gurobi and DT-STABLE.}
\label{tab:results}

\footnotesize
\renewcommand{\arraystretch}{1.05}

\resizebox{\columnwidth}{!}{%
\begin{tabular}{|c|c|c|cc|cc|cc|}
\hline
\multirow{2}{*}{\textbf{\#}} &
\multirow{2}{*}{\textbf{BS}} &
\multirow{2}{*}{\textbf{U}} &
\multicolumn{2}{c|}{\textbf{Runtime}} &
\multicolumn{2}{c|}{\textbf{SR/BS (bit/s/Hz)}} &
\multicolumn{2}{c|}{\textbf{INR Satisfaction}} \\
\cline{4-9}
& & &
\textbf{Gurobi (s)} &
\textbf{DT-STABLE (s)} &
\textbf{Gurobi} &
\textbf{DT-STABLE} &
\textbf{Gurobi} &
\textbf{DT-STABLE} \\
\hline
1  & 10 & 100 & 3.264  & 0.207 & 15.23 & 15.29 & 100\% & 75\% \\
\hline
2  & 10 & 150 & 5.395  & 0.348 & 13.81 & 14.65 & 100\% & 80\% \\
\hline
3  & 12 & 200 & 13.815 & 0.484 & 13.38 & 14.09 & 100\% & 75\% \\
\hline
4  & 12 & 220 & 15.157 & 0.543 & 13.00 & 12.60 & 100\% & 85\% \\
\hline
5  & 14 & 220 & 15.386 & 0.601 & 13.26 & 12.60 & 100\% & 70\% \\
\hline
6  & 14 & 250 & 21.214 & 0.640 & 13.13 & 12.36 & 100\% & 80\% \\
\hline
7  & 16 & 240 & 11.558 & 0.653 & 12.66 & 12.87 & 100\% & 70\% \\
\hline
8  & 16 & 270 & 14.646 & 0.697 & 12.38 & 12.59 & 100\% & 75\% \\
\hline
9  & 19 & 200 & 11.493 & 0.594 & 15.34 & 13.92 & 100\% & 70\% \\
\hline
10 & 19 & 260 & 16.861 & 0.757 & 13.30 & 14.05 & 100\% & 75\% \\
\hline
\end{tabular}%
}

\end{table}

\vspace{-0.5cm}
\section{Conclusion}
\label{sec:conclusion}
\vspace{-0.1cm}
In this paper, we introduced DT-STABLE, an O-RAN-enabled DT framework for NTN–TN spectrum coexistence. Using site-specific channel and satellite-interference maps, DT-STABLE was found to support proactive beam selection and user association without exhaustive real-time measurements. We formulated the resource allocation problem with practical constraints, showed it to be NP-hard, and developed a low-complexity Lagrangian-inspired price-based algorithm for near-real-time control. Integrated into a multi-timescale O-RAN control loop, DT-STABLE combines long-term DT mapping, mid-term corrections, and short-term scheduling.
Site-specific simulations over the 28 GHz band showed that DT-STABLE achieved an effective balance between terrestrial throughput and satellite protection. For 100 users, DT-STABLE achieves $89.63\%$ INR satisfaction and a mean sum rate of 210.96~bps/Hz, while the framework managed to attain nearly complete INR satisfaction when satellite protection was prioritized. Under dynamic blockage, DT-STABLE increased throughput by $14.8\%$ as compared to a static DT map while reducing measurement overhead by $99.71\%$. It also satisfied near-RT RIC time budget with 15.5x to 33.1x faster execution than the exact optimization. Taken together, these results demonstrate that combining site-specific DT knowledge with O-RAN control enables scalable, efficient, and interference-aware resource optimization for NTN--TN spectrum sharing.

\vspace{-0.4cm}
\appendices
\section{Derivation of DT-STABLE}
\label{app:derivation}

\subsection{Lower-Level Resource Allocation}

For a fixed interference-price vector $\boldsymbol{\lambda}^{(t)}$, the
term $\sum_m\lambda_m^{(t)}I_m^{\rm th}$ is constant with respect to
$(\mathbf{x},\mathbf{s})$. Hence, the lower-level problem is
\vspace{-1mm}
\begin{equation}
\label{eq:lower_level}
\max_{\mathbf{x},\mathbf{s}}\;
U(\mathbf{x})-\sum_{m\in\mathcal{M}}\lambda_m^{(t)}I_m(\mathbf{s}).
\vspace{-1mm}
\end{equation}

Using $K_{j,b}=\sum_u x_{u,j,b}$, the utility becomes
\vspace{-1mm}
\begin{equation}
\label{eq:utility_expand}
U(\mathbf{x})
=\sum_{u,j,b}x_{u,j,b}\log(c_{u,j,b})
-\sum_{j,b}K_{j,b}\log K_{j,b}.
\vspace{-1mm}
\end{equation}

Relaxing $x_{u,j,b},s_{j,b}\in\{0,1\}$ to $[0,1]$ yields a convex
program, since $-K\log K$ is concave and \textbf{C1}--\textbf{C3} are
affine. Dualizing $K_{j,b}=\sum_u x_{u,j,b}$ with multiplier
$\mu_{j,b}\in\mathbb{R}$ gives
\vspace{-1mm}
\begin{equation}
\label{eq:lower_lagrangian}
\begin{aligned}
\mathcal{L}
={}&\sum_{u,j,b}x_{u,j,b}
[\log(c_{u,j,b})-\mu_{j,b}]\\
&+\sum_{j,b}[\mu_{j,b}K_{j,b}-K_{j,b}\log K_{j,b}]
-\sum_{m,j,b}\lambda_m^{(t)}s_{j,b}g_{j,b,m}.
\end{aligned}
\vspace{-1mm}
\end{equation}

The stationary condition gives
$K_{j,b}^{\rm sup}=\min\{|\mathcal{U}|,\exp(\mu_{j,b}-1)\}$.
Comparing it with
$K_{j,b}^{\rm dem}=\sum_u x_{u,j,b}$ gives the load-price update
\vspace{-1mm}
\begin{equation}
\label{eq:mu_update_app}
\mu_{j,b}^{(t+1)}
=\mu_{j,b}^{(t)}
-\frac{\beta}{t+1}
\left(
K_{j,b}^{\rm sup,(t)}
-K_{j,b}^{\rm dem,(t)}
\right).
\vspace{-1mm}
\end{equation}

For fixed load prices, the association-dependent term is linear in
$\mathbf{x}_{u,\cdot}$ over the association simplex. Thus, an optimum
can be selected at a simplex vertex, yielding
\vspace{-1mm}
\begin{equation}
\label{eq:x_star}
(j_u^\star,b_u^\star)
=
\arg\max_{j,b}
\left[
\log(c_{u,j,b})-\mu_{j,b}^{(t)}
\right],
\quad
x_{u,j_u^\star,b_u^\star}^{(t)}=1.
\vspace{-1mm}
\end{equation}

The corresponding interference-adjusted beam score is
\vspace{-1mm}
\begin{equation}
\label{eq:q_beam_app}
Q_{j,b}^{(t)}
=
\sum_u x_{u,j,b}^{(t)}
[\log(c_{u,j,b})-\mu_{j,b}^{(t)}]
-\sum_m\lambda_m^{(t)}g_{j,b,m}.
\vspace{-1mm}
\end{equation}

Beams with $Q_{j,b}^{(t)}\leq0$, or with no associated UE, are
deactivated. If more than $N_{\rm RF}$ beams remain at a BS, only the
$N_{\rm RF}$ largest-score beams are retained. For the fixed tentative
association, this maximizes the retained beam-score sum subject to the
RF-chain limit. UEs on discarded beams are reassigned using
\eqref{eq:x_star} over
$\mathcal{A}^{(t)}=\{(j,b):s_{j,b}^{(t)}=1\}$, restoring
\textbf{C1}--\textbf{C3}. We emphasize that this sequential restoration is polynomial-time
but does not nessesarily guarantee global optimality, consistent with the NP-hardness
of \textbf{P0}.

\vspace{-0.4cm}

\subsection{Upper-Level Adaptive Interference Pricing}

After the lower-level allocation, the upper level computes
$v_m^{(t)}=I_m^{(t)}-I_m^{\rm th}$ and
\vspace{-1mm}
\begin{equation}
\label{eq:alpha_app}
\alpha_m^{(t)}
=
\min\!\left\{
\frac{\alpha_0}{t+1}
\left[
1+c\!\left(
e^{\gamma[v_m^{(t)}]_+}-1
\right)
\right],
\alpha_{\max}
\right\},
\vspace{-1mm}
\end{equation}
followed by
\vspace{-1mm}
\begin{equation}
\label{eq:lambda_update_app}
\lambda_m^{(t+1)}
=
\Pi_{[0,\lambda_{\max,m}]}
\!\left[
\lambda_m^{(t)}
+\alpha_m^{(t)}v_m^{(t)}
\right].
\vspace{-1mm}
\end{equation}

A positive residual increases $\lambda_m$, penalizing subsequent
interference toward satellite $m$, whereas a negative residual decreases
it. The projection keeps $\lambda_m\in[0,\lambda_{\max,m}]$. The two
levels iterate until convergence or $t=T_{\max}$, yielding a
computationally tractable but not certifiably globally optimal solution
to \textbf{P0}.

\ifCLASSOPTIONcaptionsoff
  \newpage
\fi



\bibliographystyle{IEEEtran}

\vspace{-0.3cm}
\bibliography{references}

\end{document}